\documentclass[a4paper,11pt]{article}
\usepackage{jcappub} 
\usepackage{lineno}

\usepackage{graphicx}	
\usepackage{amsmath}	
\usepackage{multicol}        
\usepackage{bm}		
\usepackage{pdflscape}	
\usepackage{fix-cm}
\usepackage{newtxtext,newtxmath}
\usepackage{stfloats}
\usepackage[T1]{fontenc}

\usepackage{caption}

\newcommand{\vk}{\mathbf k}
\newcommand{\vq}{\mathbf q}

\newcommand{\vx}{\mathbf x}

\newcommand{\drho}{\delta_{\rho}}

\newcommand{\vPsi}{\mathbf \Psi}

\newcommand{\rev}[1]{{#1}}
\newcommand{\revone}[1]{{#1}}

\title{Restoring Missing Modes of 21cm Intensity Mapping with Deep Learning: Impact on BAO Reconstruction}

\author[a]{Qian Li}
\author[a,b]{Xin Wang}
\author[a,b,c]{Xiaodong Li}
\author[d]{Jiacheng Ding}
\author[a]{Tiancheng Luan}
\author[f]{Xiaolin Luo}

\affiliation[a]{School of Physics and Astronomy, Sun Yat-Sen University, Zhuhai, 519082, China}
\affiliation[b]{CSST Science Center for the Guangdong–Hong Kong–Macau Greater Bay Area, Sun Yat-Sen University, Zhuhai, 519082, China}
\affiliation[c]{Peng Cheng Laboratory, No. 2, Xingke 1st Street, Shenzhen, 518000, China}
\affiliation[d]{Shanghai Astronomical Observatory, Chinese Academy of Sciences, No. 80 Nandan Road, Shanghai, 200030, China}
\affiliation[f]{Shanghai JiaoTong University, Minhang District, No. 800 Dongchuan Road, Shanghai, 200240, China\\}

\emailAdd{wangxin35@mail.sysu.edu.cn, lixiaod25@mail.sysu.edu.cn,luantch@mail2.sysu.edu.cn}

\abstract{In 21cm intensity mapping of the large-scale structure (LSS), regions in Fourier space could be compromised by foreground contamination. In interferometric observations, this contamination, known as the foreground wedge, is exacerbated by the chromatic response of antennas, leading to substantial data loss. Meanwhile, the baryonic acoustic oscillation (BAO) reconstruction, which operates in configuration space to "linearize" the BAO signature, offers improved constraints on the sound horizon scale. However, missing modes within these contaminated regions can negatively impact the BAO reconstruction algorithm.
To address this challenge, we employ the deep learning model U-Net to recover the lost modes before applying the BAO reconstruction algorithm. Despite hardware limitations, such as GPU memory, our results demonstrate that the AI-restored 21cm temperature map achieves a high correlation with the original signal, with a correlation ratio of approximately $0.9$ at $k\sim 1 {\rm h/Mpc}$. Furthermore,  subsequent BAO reconstruction indicates that the AI restoration has minimal impact on the performance of the `linearized' BAO signal, proving the effectiveness of the machine learning approach to mitigate the impact of foreground contamination.  Interestingly, we demonstrate that the AI model trained on coarser fields can be effectively applied to finer fields, achieving even higher correlation ratio. This success is likely attributable to the scale-invariance properties of non-linear mode coupling in large-scale structure and the hierarchical structure of the U-Net architecture. }

\begin{document}
\maketitle
\flushbottom

\section{Introduction}

In the standard cosmological model, the large-scale structure (LSS) emerges from tiny quantum fluctuations during inflation. These primordial perturbations are amplified by non-linear structure formation processes, giving rise to the cosmic web observed today. Studying such structures provides critical insights into the Universe's composition and evolutionary history \citep{colless20012df,jones20046df,drinkwater2010wigglez,anderson2014clustering,dark2016dark,collaboration2023early,ivezic2019lsst}. In addition to traditional galaxy surveys, 21cm intensity mapping (21IM) has recently emerged as a promising method for mapping LSS, by detecting the total 21cm emission of neutral hydrogen (HI) across various redshifts.  \citep{bharadwaj_using_2001,battye_neutral_2004,furlanetto_cosmology_2006,chang_intensity_2010,morales_reionization_2010,masui_measurement_2013,bull_late-time_2015,santos_cosmology_2015,villaescusa-navarro_ingredients_2018}. After the Epoch of Reionization (EoR), neutral hydrogen predominantly resides within galaxies, making 21IM an effective alternative to galaxy surveys. Unlike galaxy redshift surveys, which resolve individual galaxies, intensity mapping captures integrated light, thus offering a more efficient way to map the LSS over extremely large volumes  \citep{lidz_intensity_2011,camera_cosmology_2013,group_cosmology_2020,wang_hi_2021,schlegel_spectroscopic_2022,collaboration_euclid_2024}.

However, extracting the cosmic signal presents significant challenges due to interference from various sources, including foreground contamination from Galactic and extragalactic sources \citep{liu_data_2020,GSM2008MNRAS,GSM2017MNRAS}, radio frequency interference (RFI) \citep{Ekers_Bell_2002} and instrumental systematics \citep{morales_reionization_2010,Ding2024ApJS} etc. Among these, foregrounds are particularly problematic, as they are several orders of magnitude stronger than the expected cosmic signal \citep{bull_late-time_2015,liu_data_2020}, significantly hindering progress in measuring the auto-correlation of the signal. Additionally, the chromatic response of the instrument can further complicate situations by introducing extra coupling that contaminates a large region in Fourier space, known as the foreground wedge \citep{morales_reionization_2010}. 
To address these issues, many techniques have been developed, leveraging the spectral smoothness \citep{Santos2005ApJ,Liu2009MNRAS}, the dominance \citep{masui_measurement_2013,switzer2013determination,switzer_interpreting_2015,bigot2015simulations}, or statistical properties \citep{chapman2012foreground,wolz2014effect} of the foreground. An alternative approach is to ignore the foreground-contaminated regions and conduct cosmological measurements using only the remaining data, a method known as foreground avoidance. However, in either case, a significant portion of Fourier space is compromised or lost by foreground interference.

\rev{This missing information} can significantly impact cosmological measurements. For one, the loss of Fourier modes reduces the statistical constraining power.
For baryonic acoustic oscillations (BAO), this loss also affects data post-processing. Most current BAO measurements in galaxy surveys utilize a technique known as BAO reconstruction \citep{eisenstein_improving_2007,ZYP17a,Yu2017a,Wang17,SBZ17,obuljen_baryon_2017,Wang_2019,BJLi2019MNRAS}, which aims to mitigate the non-linear degradation of the BAO signal. Despite the technical differences among various reconstruction algorithms, these methods generally seek to solve the continuity equation to `linearize' the BAO signature. \rev{However, contamination of a large fraction of Fourier modes can significantly impact the reconstruction outcome.  \cite{seo_foreground_2016} investigated this effect using Fisher information at redshifts $z=1-2$ and found that it could increase errors by a factor of $3-4$ for the angular diameter distance and by $\sim 1.5$ for $H(z)$. One potential solution is to use a low-density galaxy sample to compensate for the affected modes  \citep{cohn_combining_2016}. 
Alternatively, information loss can be mitigated by leveraging nonlinear mode coupling. For example, the tidal reconstruction method \citep{zhu2016} employs a quadratic estimator derived from the tidal field to recover large-scale modes from small-scale perturbations. However, as a perturbative approach, its effectiveness is primarily limited to large scales. }

With the widespread application of artificial intelligence (AI), machine learning (ML) has become a powerful tool for processing large-scale structure data and \rev{analyzing cosmic datasets} \citep{villaescusa-navarro_weighing_2015,sadeh_annz2_2016,ho_robust_2019,agarap_deep_2019,wu_cosmic_2021,villanueva-domingo_inferring_2022,schaurecker_super-resolving_2022,shi_21_2024,masipa_emulating_2023}. It has also proven to be highly effective in addressing high-dimensional astronomical problems, such as reconstructing BAO signals from dark matter density fields \citep{mao_baryon_2020}, isolating the EoR signals with complex frequency-dependent beam effects \citep{li_separating_2019}, and identifying ionization regions in images with instrument noise \citep{gagnon-hartman_recovering_2021,bianco_deep_2021}.
Therefore, in this study, we aims to apply the deep learning technique, particularly the U-Net model, to recover the large-scale structure modes in Fourier regions affected by foreground contamination. Additionally, we will examine the impact of the AI-restored data on BAO reconstruction. 
The structure of this paper is as follows: Section \ref{sec:introfg} provides a brief overview of foreground contamination and its impact on BAO reconstruction. In section \ref{sec:method}, we introduce our main methodology, including details of the simulation data for 21cm intensity mapping, the U-Net architecture, and the BAO reconstruction process. The results are presented in Section \ref{sec:result}, and we conclude with a discussion in Section \ref{sec:conclusion}.

\section{Foreground Contamination and the Impact on BAO Reconstruction}
\label{sec:introfg}

\label{sec:maths} 
In both single-dish and interferometric observations, the spectral smoothness of the foreground indicates that the LSS signal may be compromised within the Fourier region where the wavenumbers along the line-of-sight direction are small, i.e. $k_{\parallel}<k_{\rm int}$ where  $k_{\parallel}$ represent the line-of-sight component of the wave vector $\vk$. \rev{This is commonly} referred to as the intrinsic foreground. The exact value of $k_{\rm int}$ depends on the specific foreground removal techniques employed and specific instrumental systematics; for our analysis, we have adopted a reasonable value of $k_{\rm int} = 0.1~{\rm h/Mpc}$.
Furthermore, the chromatic response of \rev{the} interferometer complicates the situation \citep{datta_bright_2010,morales_four_2012,parsons_per-baseline_2012} by introducing additional coupling between spectral and spatial information. This coupling can lead to foreground leakage from low $k_\parallel$ to high $k_\perp$, with the effect being particularly severe at high $k_\perp$, where $k _{\perp} $ is the transverse components of $k$.

As extensively discussed in the literature, the mixing induced by instruments presents both challenges and opportunities. Assuming a simple toy model for the foreground, where the power spectrum $P^{\rm fg}(k)$ is proportional to the Dirac delta function $\delta^D(k_{\parallel})$ along LoS, the observed power spectrum can be approximated as \citep{liu_data_2020}
\begin{eqnarray}
    \hat{P}(\vk) \propto \bar{A}_p^2 \left [ \frac{k_{\parallel}}{k_{\perp}} \frac{c}{H_0} \frac{(1+z)}{E(z)}  \right]
\end{eqnarray}
where $\bar{A}_p^2$ represents the average squared primary beam profile, integrated across directions orthogonal to the baseline. The function $E(z)$ is defined as 
\begin{eqnarray}
     E(z) \equiv \sqrt{\Omega_m (1+z)^3 + \Omega_k (1+z)^2 + \Omega_\Lambda}
	\label{eq:Ez}
\end{eqnarray}
with $\Omega_m,\Omega_\Lambda$ and $\Omega_k$ are the density parameter of matter, cosmological constant and curvature respectively. 
Consequently, a foreground confined in low wavenumber along LoS $k_{\parallel}$ can significantly spread into high $k_{\perp}$ values. However, since the primary beam of an antenna typically diminishes at the edge, it naturally sets a limit on this foreground mixing. If we assume the primary beam attenuates to zero at an angle $\theta_{\rm FOV}$, the relationship delineating the foreground wedge can be described as 
\begin{eqnarray}
    k_{\parallel} \leq k_{\perp} \frac{E(z)}{(1+z)}  \theta_{\rm FOV} 
    \int_{0}^{z} \frac{dz'}{E(z')} . 
	\label{eq:kp}
\end{eqnarray}
in the two-dimensional $k_{\parallel}-k_{\perp}$ plane. In this paper, \rev{we consider $\theta_{\rm FOV}=\pi/2$}, at the redshift of $z=1$, and \rev{the} cosmological parameters as follows $\Omega_m=0.31,\Omega_b=0.048,\Omega_{\Lambda}=0.69,\sigma_8=0.82,h=0.68$.
The combination of intrinsic foreground and the mode mixing wedge results in significant information loss, which, as will be discussed, substantially impacts BAO reconstruction.


Despite its robustness and enormous success in constraining cosmological parameters in LSS surveys, the baryonic acoustic oscillation is still subject to various systematics. One of the most well-understood effects is the broadening of the BAO peak caused by the non-linear evolution of the bulk velocity of LSS \citep{BAO2007ApJEis}. \rev{In  Fourier space}, this manifests as a dampening of the BAO wiggles, which can be modeled by a simple exponential function. Compared to the linear BAO peak, this effect reduces \rev{the accuracy of measurement in a given LSS survey when extracting the sound horizon scale}. Moreover, the same non-linear process also leads to the emergence of non-Gaussianity, which results in additional information leakage into \rev{higher-order statistics} and further decreases the overall constraining power of the measurements. 
Various techniques have been developed to reverse such process. For example, \cite{eisenstein_improving_2007} proposed a reconstruction algorithm based on the linear Zel'dovich approximation which can partially reverse nonlinear degradation of BAO signal. This technique has been widely used in recent galaxy surveys \citep{SDSSrec2012MNRAS,SDSS3rec2016MNRAS,SDSS4rec2018ApJ,DESIBAO2024arXiv}. \rev{Additionally, \cite{obuljen_baryon_2017} extended this method to pixelated maps for applications in 21cm intensity mapping. }

Recently, many other nonlinear reconstruction methods have been proposed \citep{ZYP17a,Yu2017a,Wang17,SBZ17,Wang_2019,BJLi2019MNRAS}, setting aside their technical specifics, all of these algorithms aim to solve the continuity equation
\begin{eqnarray}
\label{eqn:contn}
 \det \left ( \frac{ \partial x_i}{\partial q_j }\right) = \det\left ( \delta^K_{ij} + \partial^2_{ij} \phi \right) = 
 \frac{\rho_{\rm init}}{\rho }  = \frac{1}{1+\drho}, 
\end{eqnarray}
where $\vq$ and $\vx$ are Lagrangian and Eulerian coordinates of particles respectively, $\delta^K_{ij}$ is the Kronecker delta, the displacement potential and vector are defined with $\vPsi(\vq) = \vx - \vq = \nabla \phi (\vq) $. By solving equation (\ref{eqn:contn}), \rev{the reconstruction algorithm reverses such a non-linear transformation} and produces a reconstructed $\phi_{\rm rec}$ field on a grid that is close to the Lagrangian coordinates. In the following, we will denote the Laplacian of the reconstructed field $\delta_{\rm rec} = \nabla^2 \phi_{\rm rec}$. Consequently, the BAO signature of the reconstructed field closely \rev{resembles} that of the linear theory. 
Since these algorithms solve for the displacement field in the configuration space, in 21cm intensity mapping experiments, the missing modes due to the foreground contamination will inevitably impact the performance of the reconstruction.

\section{METHOD}
\label{sec:method}

\subsection{Data Preparation}
\label{sec:dmsim}

\rev{At low redshifts, neutral hydrogen primarily resides within galaxies. In this subsection, we describe the generation of training data for our AI model. 
Given the computational cost of producing large simulation datasets, we adopt several simplifications. First, We use the COLA (COmoving Lagrangian Acceleration) simulations code \footnote{https://bitbucket.org/tassev/colacode/src/hg/} \citep{COLA2013JCAP}, which integrate second-order Lagrangian Perturbation Theory (2LPT) with N-body algorithms, to rapidly simulate multiple realizations of dark matter distributions. Second, to assign HI mass $M_\mathrm{HI}$, we follow the fitting recipe from \cite{villaescusa-navarro_ingredients_2018} to populate dark matter halos with neutral hydrogen. Due to the limited mass resolution, we also introduce a subgrid population to account for halos below the simulation’s resolution threshold. 
}

For the dark matter simulation, \rev{we conducted 37 realizations} using the same set of cosmological parameters described previously but with varying initial seeds. Each simulation contains $1024^3$ particles within a cubic box measuring $800 {\rm h^{-1} Mpc}$. The initial conditions for these simulations were set at a redshift of $z_{\rm init} = 29$.
We then generate the corresponding distributions of neutral hydrogen for each realization. Given the relatively low mass resolution of our dark matter simulations, the resolved halo masses are significantly larger than the minimum halo mass that can physically host HI. Consequently, our HI map consists of two distinct components: resolved halos and a subgrid contribution.
For resolved halos, we employ the Robust Overdensity Calculation using K-Space Topologically Adaptive Refinement (ROCKSTAR) halo finder \citep{behroozi_rockstar_2013} to identify halo catalogs. This includes determining their locations, masses, and peculiar velocities. Furthermore, we account for the redshift space distortions (RSD) by adjusting the real space coordinates of halos according to their peculiar velocities $v_r = \textbf{v} \cdot \hat{\textbf{r}}$, along the line-of-sight (LoS) direction, which we designate as the $z$ axis.
The redshift space position $\textbf{s}$ is then defined by
\begin{eqnarray}
\label{eqn:rsd}
     \textbf{s} = \textbf{x} + \frac{\textbf{v} \cdot \hat{\textbf{z}}} {aH(z)} \hat{\textbf{z}} 
\label{eq:rsd}
\end{eqnarray}
Here, $\textbf{x}$ represents the real space comoving position, and $\textbf{v} \cdot \hat{\textbf{z}}$ is the peculiar velocity along LoS. Additionally, $a$ is the scale factor, $H(z) \equiv H_0 E(z)$  is the Hubble parameter at redshift $z$. These halos will subsequently be populated with HI using a halo mass relation, as discussed in the next subsection.

For subgrid contribution, \rev{we assume that} these low-mass halos trace the dark matter field. Therefore, with the position and the peculiar velocity of \rev{the dark matter particles}, we construct the \rev{redshift space} matter distribution using Equation (\ref{eqn:rsd}). 
In each cell with volume $V_{\rm cell}$ and mass $M_{\rm cell}$ at redshift $z$, \rev{we assume that}  the number of halos with mass $m$ follows a Poisson distribution, with the average number density determined by the halo mass function.  For halos that are viralized at redshift $z'$, the average number density can be calculated by \citep{CoorySheth2002} 
\begin{eqnarray}
   n(m, z' \mid M_{\rm{cell}}, z) = f(\nu_{10})\frac{\Omega_m \rho_{\mathrm{crit}}}{m}\frac{d\nu_{10}}{dm}, 
\end{eqnarray}
with 
\begin{eqnarray}
    \nu_{10}=\frac{\left[\delta_{sc}(z')-\delta_0(\delta,z)\right]^2}{\sigma^2(m)-\sigma^2(M_{\mathrm{cell}})}. 
\end{eqnarray}
Here $\rho_\mathrm{crit}$ is the critical density, $\sigma(m)$ represents the variance of density fluctuations on mass $m$ and $\delta_{\text{sc}}(z')$ is the critical density for spherical collapse at redshift $z'$ extrapolated to the present using the linear theory. The term $\delta_0(\delta, z)$ corresponds to the initial density necessary for a region to reach $\delta$ at redshift $z$. 
Therefore, within each cell of volume $V_\mathrm{cell}$, we generate a random sample of low-mass halos within a specific mass bin, \rev{ensuring that} $n_\mathrm{cell}(m)$ follows a Poisson distribution with mean $n(m)$. Similar to resolved halos, these randomly generated halos are then converted into HI, and subsequently, into brightness temperature.

Following the recipe described in \cite{villaescusa-navarro_ingredients_2018}, the HI mass within a halo with mass $m$ at redshift $z$ is calculated using the fitting formula
\begin{eqnarray}
    M_{\rm HI}(m)=M_0 \left(\frac{m}{M_{\rm min}} \right)^\alpha \exp{\left(-\left(\frac{M_{\rm min}}{m}\right)^{0.35}\right)}
	\label{eq:halomf}
\end{eqnarray}
Here, $M_0, M_{\rm min}, \alpha$ are redshift-dependent parameters that interpolated from the fitted values provided in \cite{villaescusa-navarro_ingredients_2018}. 
The brightness temperature at redshift $z$ is subsequently calculated as 
\begin{eqnarray}
    T_b(z)=23 (1+\delta_\mathrm{b} ) \left[ \left(\frac{0.15}{\Omega_m}\right) \left(\frac{1+z}{10} \right)  \right]   ^{0.5}\left (\frac{\Omega_b h}{0.02} \right)  {\rm mK}
	\label{eq:Tb}
\end{eqnarray}
where $\delta_\mathrm{b}$ is the baryon fluctuation. Since after the EoR, HI mainly resides within galaxies, we assume that the HI fraction $x_{\mathrm{HI}}$ is unity and that the spin temperature is significantly higher than the cosmic microwave background radiation temperature $T_\mathrm{S} \gg T_\gamma$. The baryon fluctuation  $(1+\delta_\mathrm{b})$ in each cell is related to the HI mass by the equation
\begin{eqnarray}
    1+\delta_\mathrm{b} (z)= \frac{M_{\rm HI}(m,z)}{V_{\rm cell}  f_\mathrm{H}  \Omega_{\mathrm{b}}\rho_\mathrm{crit}}
	\label{eq:fracHI}
\end{eqnarray}
where $f_{\mathrm{H}} = 0.76$ is the hydrogen fraction. 
After generating these observed brightness temperature data, we further perform data augmentation by \rev{rotating the datacube around a fixed line-of-sight direction as the rotation axis to generate additional samples. 
This process results in a total of $37\times 4 =148$ samples of 21cm brightness temperature.} The dataset is then split into three subsets: $80\%$ is used for training the U-Net models, $10\%$ is reserved for validation to fine-tune hyperparameters, and the remaining $10\%$ is used for testing to evaluate the model's generalization and performance.

\rev{However, our HI modeling has clear limitations. The empirical HI assignment does not fully capture its true clustering properties, which may impact our AI model by altering the underlying mode coupling. Additionally, the low mass resolution of our dark matter simulation requires subgrid modeling of HI content, introducing biases and failing to preserve the spatial correlations present in a realistic HI distribution. Furthermore, \cite{villaescusa-navarro_ingredients_2018} showed that while the bulk velocity of HI closely follows that of its host halo in smaller halos, it deviates significantly for larger halos. Moreover, HI velocity dispersion is generally lower than that of dark matter, particularly in small halos. As a result, the redshift-space distortions in both resolved and unresolved halos in our model may be biased. 
Despite these drawbacks, this method remains a practical compromise for generating the necessary training data. In future work, we will focus on higher-resolution simulations and explore alternative HI modeling and velocity treatments to improve accuracy. 
}

\begin{figure*}
    \includegraphics[width=1.05\linewidth]{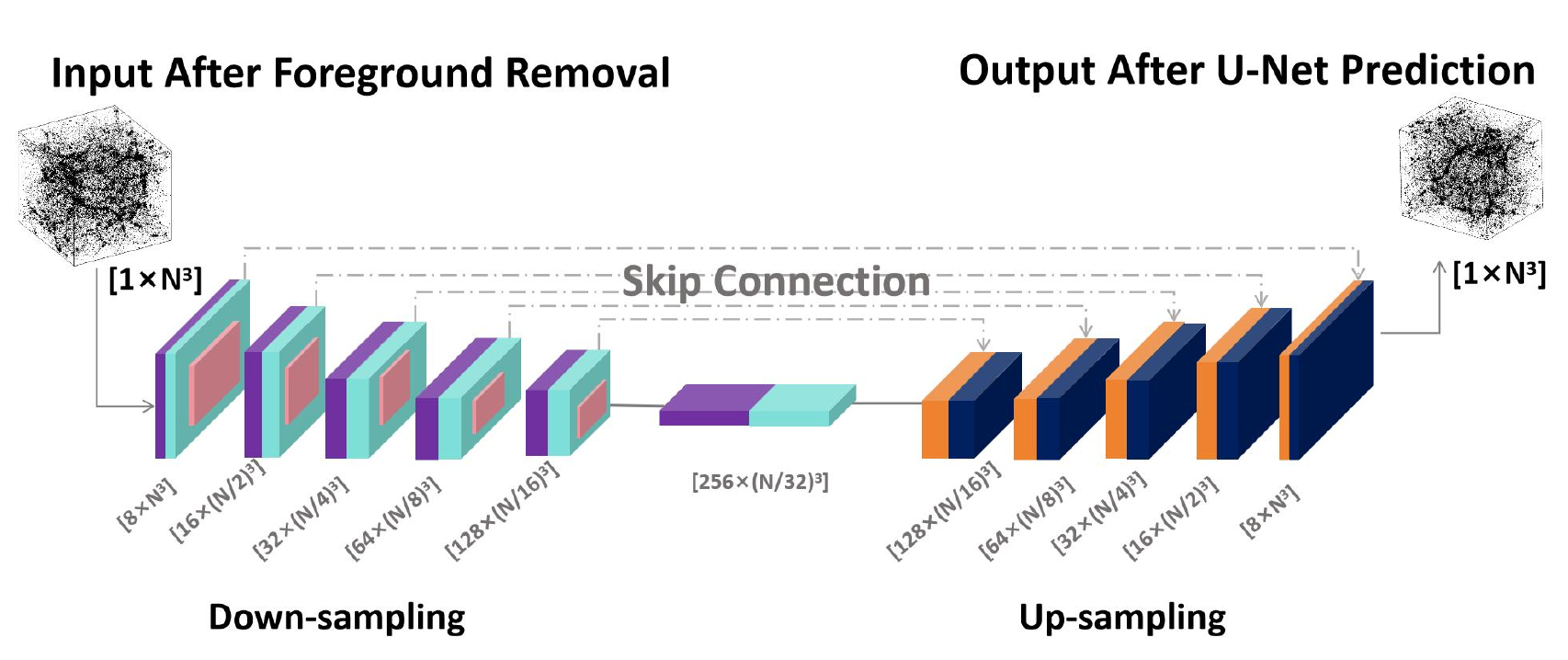}
    \caption{     
The block architecture diagram of the U-Net model. In our U-Net model, each level of the down-sampling phase includes pairs of convolution layers with $3 \times 3 \times 3$ kernels, followed by a  $2 \times 2 \times 2$ max pooling layer to reduce spatial dimensions. Each convolution layer is followed by a ReLU activation function to introduce non-linearity. During the up-sampling phase, transposed convolutions are utilized to restore the spatial resolution lost in the down-sampling. Additionally, skip connections link corresponding layers from the down-sampling phase to the up-sampling phase, enhancing the network's ability to capture fine-grained details.
  }
\label{fig:unet}
\end{figure*}

\subsection{U-Net Architecture and Training}
\label{sec:Unetarch}
To address the issue of missing modes, we employ deep learning techniques, specifically the U-Net architecture, for their recovery. U-Net is a widely adopted network for image segmentation that is structured around a fully convolutional neural network \citep{ronneberger_u-net_2015,shelhamer_fully_2017}.  While the ability to recover these missing modes is due to the non-linear mode coupling in LSS, which is more conveniently described in Fourier space. However, such coupling in Fourier space is intrinsically non-local and often involves features that are quite distant from each other. Meanwhile, standard hierarchical convolutional networks, which progressively connect information from nearby pixels, struggle to effectively capture such non-local features. In contrast, the same effect is much more localized in configuration space, thereby making it a more suitable domain for applying the neural network.

As shown in Figure \ref{fig:unet}, our U-Net model consists of five layers each of \rev{the} down-sampling and up-sampling process. In the down-sampling phase, the encoder incorporates convolution and pooling layers to gradually reduce the high spatial resolution of the simulated data. To process large volumetric data, we use successive double convolutional layers of $3 \times 3 \times 3 $ followed by a max pooling layer with $2 \times 2 \times 2 $ kernel size to efficiently \rev{capture three-dimensional spatial features}.  In the up-sampling phase, transposed convolution layers of $2 \times 2 \times 2$ are utilized to restore spatial resolutions. Additionally, skip connections create direct links between down-sampling and up-sampling layers with matching corresponding resolutions, allowing easier flow of features and gradients during backpropagation, while enhancing the preservation of complex details and helping to alleviate computational overhead. 
The ReLU activation function is applied after each convolution layer, allowing the network to capture and retain effective feature information from deeper levels.


The training of neural networks begins with forward propagation, during which the network processes input data to extract features and produce predictions. \rev{Therefore,} the quality of this input data significantly impacts \rev{both} the network’s training efficacy and the accuracy of its predictions. 
Training neural networks requires considerable computational resources to update the weights and biases during training. Higher-resolution input data allows the model to capture finer details of LSS, which encode valuable information about non-linear mode coupling. However, this improvement comes at the cost of increased computational demands, as higher resolution results in larger input volumes and a corresponding increase in the number of hyperparameters that must be processed.

To optimize model performance while managing computational constraints, we divided the extensive dataset into smaller batches and experimented with different resolutions to evaluate their effects on model efficacy. Our goal is to achieve an optimal balance between computational limitations and the maximization of usable data. From the previously described simulated 21cm brightness temperature data, we randomly selected $80$ percent as the training set, in which foreground contaminated modes have been excluded. 
\revone{After some testing, we adopted a two-step training procedure to maximize performance within our GPU constraints, using the same U-Net architecture (Figure \ref{fig:unet}). In the first stage, we trained the AI model using simulated data cubes with a grid size of $128^3$ (stage-one model). This initial model already performed reasonably well, effectively capturing mode-coupling features and demonstrating strong capability in restoring high-quality modes, even when applied to higher-resolution data. However, it exhibited noticeable amplitude deviations in the deeply nonlinear regime for higher resolution maps.
To further improve performance, we retrained this model using a smaller dataset of only $13$ simulated fields at a higher resolution of $256^3$ (stage-two model), again limited by available GPU resources. 
}

During training, we compute the Mean Squared Error (MSE) loss function, defined as the average squared difference between the predicted field $T^{\rm pred}$ and actual field $T^{\rm true}$, i.e. 
\begin{eqnarray}
 \textsc{Loss} = \frac{1}{N} \sum\limits_{i=1}^N \left (T^{\rm pred}_i - T^{\rm true}_{i} \right)^2
\end{eqnarray}
where $N$ represents the number of training samples in a single batch. This loss function quantifies the deviation of the model’s predictions from the true values for each batch. With \rev{the} above loss function, we employ the Adam optimizer \citep{kingma_adam_2017} to adjust the model weights and biases by minimizing the squared errors during the training loop, with a learning rate of $10^{-3}$.
The backpropagation process is initiated to fine-tune the network parameters, allowing the generated results to progressively converge with the true values. This iterative adjustment continues until the network achieves convergence.

\subsection{BAO Reconstruction}

After restoring the modes using machine learning, we proceeded to assess their impact on the non-linear iterative BAO reconstruction. Various algorithms are available for this purpose. In this study, we particularly adopted a particle-based reconstruction algorithm, similar to the one described in \cite{Schmittfull_2017PhRvD} for its simplicity and computational efficiency. A detailed examination reveals that the reconstruction performance of this code is very similar or even slightly better than the field-based algorithm we have developed previously \citep{ZYP17a,Wang17}. 
Both algorithms solve the continuity equation (\ref{eqn:contn}) for the non-linear displacement field $\vPsi$. In the particle-based approach, this is achieved by iteratively applying the Zel'dovich approximation to a field that is smoothed with gradually smaller scales. 
Therefore, the algorithm executes the following steps during each $i$-th iteration:  
\begin{itemize}
    \item Calculate the density map $\rho_{i}$ from \rev{the} distribution of particles, using \rev{the} Cloud-in-Cloud algorithm (CIC).
    \item Apply a Gaussian smoothing window $W_{\rm G}=\exp\left(-k^2 R_i^2/2\right)$ to the density map $\rho_{i}(\vx)$.
    \item Solve for the linearized displacement potential $\psi_i(\vk)=-\rho_{i}(\vk)/k^2$, and then convert it back to the configuration space. 
    \item Move all particles backward by applying the negative of the displacement, $- \nabla \psi_i$. 
\end{itemize}
Here, we set the smoothing scale for the $i$-th iteration as $R_i=\left( \sqrt{2}/2\right)^{i-1} R_{\rm ini}$, where $R_{\rm ini} = 15~ {\rm Mpc/h}$. After approximately $9$ iterations, the smoothing scale reduces to about $R\sim 1$, at which point the algorithm exists the iteration. We then compute the combined non-linear displacement $\vPsi_{\rm rec}= \sum_i \nabla \psi_i $. The reconstructed scalar field is then defined as the divergence of the displacement, $\delta_{\rm rec} = \nabla \cdot \vPsi_{\rm rec}$.

Since the product of a 21cm intensity mapping experiment is a temperature map, it is necessary to convert the temperature fluctuation into distribution of particles. To achieve this, we employ supersampling of the density field to generate particles. To maintain high resolution in the particle representation and ensure the smoothness of the field, the number of particles, $N_{\rm part}$, is set to be at least $2^3$ times greater than the number of pixels in the original field.  
Furthermore, since the reconstruction algorithm does not differentiate between the dark matter field and any biased tracers, it is crucial to de-bias the temperature distribution so that the input map provided to the algorithm accurately represents an estimation of the underlying dark matter distribution. For such purpose, we define the input density distribution 
\begin{eqnarray}
\label{eqn:debias}
    \hat{\delta}_\mathrm{input} (\vx) = \delta T (\vx) /
    \left( \bar{T}_\mathrm{HI} b_\mathrm{T} \right). 
\end{eqnarray}
\rev{Here, $\bar{T}_\mathrm{HI}$ is the average HI temperature, $b_{T}$ is the linear bias coefficient, estimated from the cross-correlation between dark matter field and the temperature fluctuation map. Specifically, we averaged the scale-dependent $\bar{T}_\mathrm{HI} b_\mathrm{T} (k) = \langle \delta_\mathrm{dm} \delta T \rangle / \langle  \delta_\mathrm{dm}  \delta_\mathrm{dm} \rangle$ up to a cut-off scale $k_\mathrm{bcut}=0.035 ~ {\rm h/Mpc}$. 
In practice, without direct access to the dark matter distribution, the factor $\bar{T}_\mathrm{HI} b_\mathrm{T}$ must be measured using alternative modeling methods, which can introduce additional errors. 
Fortunately, as proposed by \cite{Wang_2019}, one can iteratively reconstruct the field with different bias factors (or the combined term $\bar{T}_\mathrm{HI} b_\mathrm{T}$ here) and minimize BAO smearing by simultaneously using both the peak location and width of the BAO signature. However, this approach is beyond the scope of the current paper and will be explored in future work. 
}

\begin{figure*}
    \centering
\includegraphics[width=\columnwidth]{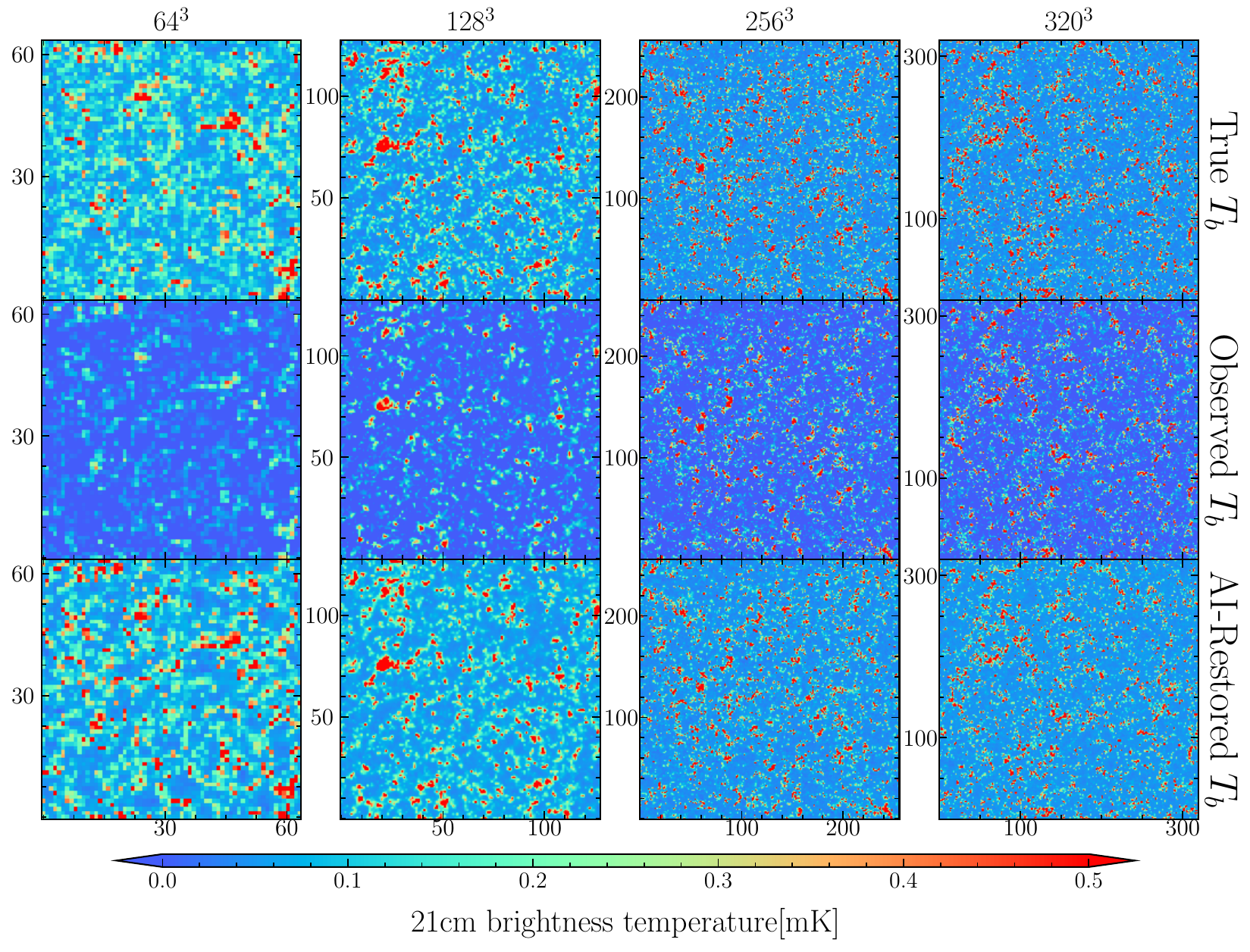}
    \caption{Visual comparison between the true brightness temperature ({\it upper} panels), the observed temperature with intrinsic and foreground wedge removed ({\it middle} panels), and the U-Net restored temperature ({\it lower} panels). Different columns demonstrate maps at various resolutions, ranging from $64^3$ ({\it first} column) to $128^3$ ({\it second}), $256^3$ ({\it third}), and $320^3$ ({\it last} column).  \revone{Here, our low resolution results ($64^3$ and $128^3$) were obtained using the stage-one AI model trained at a resolution of $128^3$, while the higher-resolution results ($256^3$ and $320^3$) were produced using the stage-two model, which was retrained with $256^3$ data. (The same applies to the results in the following.)}  
    All brightness temperatures are in unit of ${\rm mK}$. 
    }
    \label{fig:multimK}
\end{figure*}

 \begin{figure*}
    \centering
    \includegraphics[width=\columnwidth]{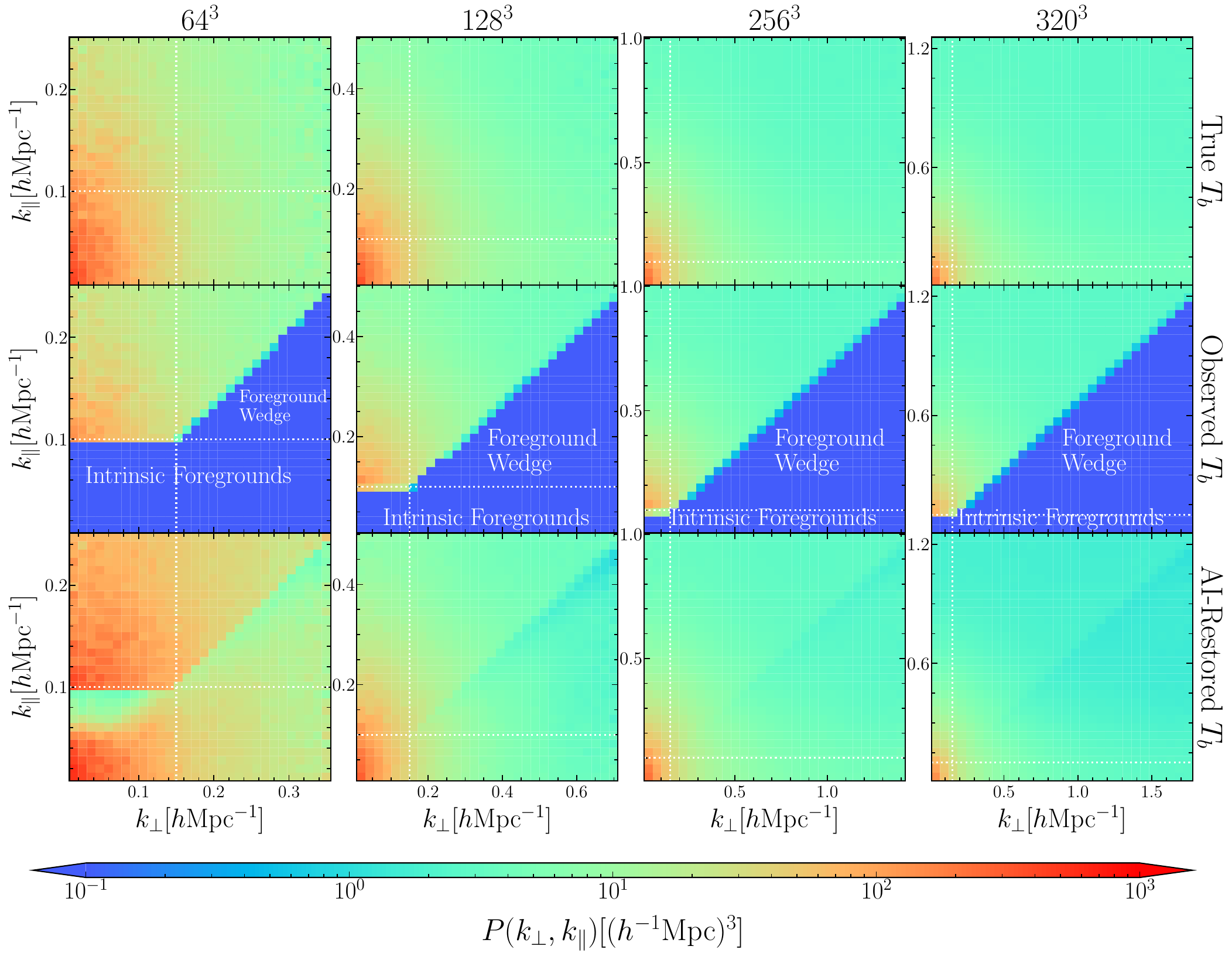}
    \caption{Comparison of the two-dimensional power spectrum $P(k_{\perp}, k_{\parallel})$ of 21cm brightness temperature across various fields and resolutions. The first row displays the power spectrum of the true brightness temperature. The second row presents the observed map, with modes contaminated by foregrounds removed, specifically those with $k _{\parallel}<0.1 ~{\rm h/Mpc}$ and within the wedge. The last row displays the power restored by the U-Net model. Additionally, different columns correspond to power spectra calculated from maps at varying resolutions, starting from $64^3$ up to $320^3$. 
    }
    \label{fig:multipk2d}
\end{figure*}

\subsection{BAO Extraction and Parametric Fitting}
\label{sec:paramfit}
\rev{After reconstruction, we examine the BAO signature of the processed power spectrum to assess the reconstruction performance. A commonly used approach is to normalize by the power spectrum of seed-matched no-wiggle simulations, allowing the BAO signature to be defined as 
\begin{eqnarray}
\label{eqn:BAOsim}
    S^\mathrm{sim}_\mathrm{BAO} (k) = \frac{P_\mathrm{BAO}-P_\mathrm{nwig}}{P_\mathrm{nwig}} (k), 
	\label{eq:bao_s}
\end{eqnarray}
where $P_{\rm BAO}$ represents the power spectrum with BAO, and $P_{\rm nwig}$ denotes the corresponding `no-wiggle' power spectrum, and the superscript $\mathrm{sim}$ indicates that this BAO signal is derived directly from simulation-based manipulation. 
\revone{To examine the performance of our AI model under the most idealized conditions, we first process both $P_\mathrm{BAO}$ and its corresponding $P_\mathrm{nwig}$ through our AI restoration and BAO reconstruction pipeline. This allows us to establish a baseline for evaluating the performance of the combined AI restoration and BAO reconstruction procedure.}

Of course, in real observations, one does not have access to a Universe with a no-wiggle power spectrum. \revone{Since various effects can distort the overall shape of the processed power spectrum, evaluating the performance of BAO reconstruction becomes challenging. Therefore, alternative methods are required to robustly extract the BAO signal. Consequently, we adopt the standard BAO measurement approach \citep{SDSS3BAO_2012MNRAS}, fitting the observationally processed power spectrum with a template that accounts for the BAO signature, the overall broadband shape, and additional distortions introduced by processing pipeline. Specifically, we model the template power spectrum as 
\begin{eqnarray}
\label{eqn:fittemp}
P^\mathrm{temp}(k) = P^\mathrm{temp}_\mathrm{nwig}(k) T(k) 
 \left [ S^\mathrm{temp}_\mathrm{BAO}(k) + 1 \right]. 
\end{eqnarray}
Here, $P^\mathrm{temp}_\mathrm{nwig}(k)$ is the no-wiggle power spectrum template, unaffected by foreground removal or AI processing}, $T(k)$ is a fourth-order polynomial 
\begin{eqnarray}
   T(k)=\sum_{i=0}^{4} a_i k^i , 
\end{eqnarray}
accounting for broadband shape distortion. \revone{And finally, $S^\mathrm{temp}_\mathrm{BAO}(k)$ is the theoretical BAO template defined as}
\begin{eqnarray}
\label{eqn:BAOsig}
S^\mathrm{temp}_\mathrm{BAO}(k) = S^\mathrm{temp}_{\mathrm{BAO, lin}}(k) \exp \left[ -k^2 \Sigma^2 /2\right].
\end{eqnarray}
\revone{Here, $S^\mathrm{temp}_{\mathrm{BAO, lin}}(k)$ is the theoretical linear BAO signal obtained by dividing the output of the Boltzmann code \texttt{CAMB} \citep{2011ascl.soft02026L} by a smooth, fitted no-wiggle power spectrum. The parameter $\Sigma$ quantifies the degree of nonlinear smearing or the effect of reconstruction. }
\revone{Therefore, compared with Equation (\ref{eqn:BAOsim}), this approach follow a more observationally motivated procedure and avoids relying on AI- and BAO-reconstructed no-wiggle simulations. All AI-induced shape distortions in the band power are absorbed into the transfer function $T(k)$, ensuring an unbiased evaluation of the process. Meanwhile, the effectiveness of the BAO reconstruction—with or without AI restoration—is reflected in the fitted damping parameter $\Sigma$. As we will show, improved BAO reconstruction results in a smaller $\Sigma$, indicating reduced nonlinear smearing.}

Notice that since we are focusing on evaluating the AI restoration in the context of BAO reconstruction, we do not include cosmology dependent scaling factors that account for the geometric shift of the wavenumber $k$. 
Additionally, for simplicity, we use only the diagonal components of the covariance matrix in the $\chi^2$ estimation, which was computed using our original $37$ simulated training samples.
}

\section{RESULT}
\label{sec:result}

\rev{}

\rev{\subsection{AI-restored Maps}}

In Figure \ref{fig:multimK}, we present a visual comparison of true temperature maps ({\it top} panels), observed maps with foreground-contaminated modes removed ({\it middle} panels), and \rev{the AI-restored} map ({\it lower} panels). Different columns display the same map at varying grid sizes, ranging from $64^3$, to $128^3$, $256^3$, and $320^3$. \revone{Due to a slight performance drop of the stage-two retrained AI model on lower-resolution maps, the $64^3$ and $128^3$ results were obtained using the stage-one AI model trained at $128^3$, while the higher-resolution results ($256^3$ and $320^3$) were generated using the stage-two model retrained with $256^3$ data.}  
As demonstrated, in all these cases, the AI-restored maps successfully resemble the true brightness temperature maps in each column.

\begin{figure*}
    \centering
    \includegraphics[width=1.0\linewidth]{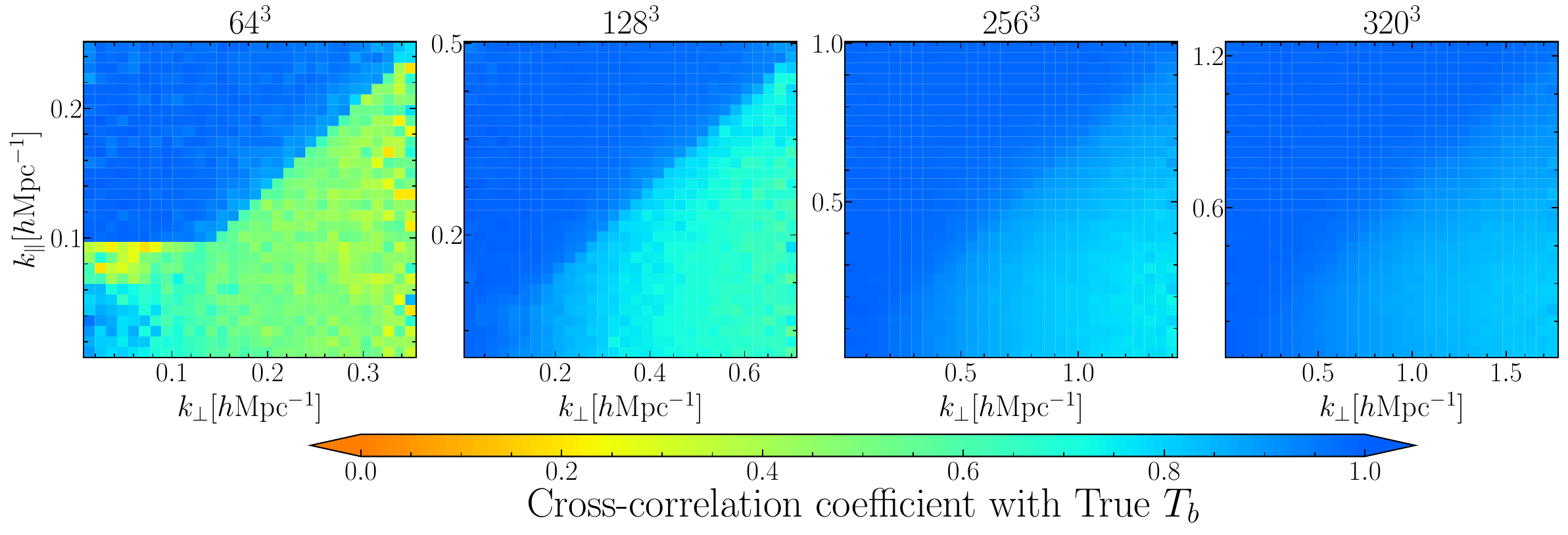}
    \caption{The two-dimensional cross-correlation ratio between the AI restored fields and the true temperature maps across varying resolutions. From left to right, resolutions of $64^3$, to $128^3$, $256^3$ and $320^3$ are displayed, demonstrating that applying the model to higher resolutions  significantly improve the cross-correlation ratio. }
    \label{fig:multicof}
\end{figure*}

\revone{Despite variations in performance when applying our AI model to maps of different resolutions, it is noteworthy that it consistently succeeds in recovering the missing modes.
Before delving into the detailed performance analysis in each case, let us first explore the underlying reasons behind this apparent scale invariance of the AI model.} This property can largely be attributed to two factors: the self-similarity exhibited by \rev{the evolution of large-scale structure} and the hierarchical structure of \rev{the U-Net model}. Specifically, from the perspective of perturbation theory, the mode-coupling coefficients appeared in the non-linear dynamical evolution of the perturbation \citep{SPTreview2002PhR}
\begin{eqnarray}
\alpha (\vk_1, \vk_2) &=& (\vk_1+\vk_2)\cdot \vk_1 /k_1^2  \nonumber \\
\beta(\vk_1, \vk_2) &=& \left| \vk_1+\vk_2\right|^2 \left[\vk_1 \cdot \vk_2 / \left (2 k_1^2 k_2^2 \right)\right], 
\end{eqnarray}
is deterministic and predictable. Therefore, the patterns learned by the deep learning model at coarser grid resolution are not only meaningful but also representative of physical processes that persist across scales. 
On the other hand, the architecture of U-Net, with its hierarchical structure of convolutional layers designed to capture features at multiple scales, is particularly well-suited to leveraging such scale invariance of the large-scale structure data.

To evaluate the performance of our U-Net model, we present the two-dimensional power spectrum $P(k_{\perp}, k_{\parallel})$ of various fields in Figure \ref{fig:multipk2d}. The upper panels show the power spectrum of the true brightness temperature, the middle panels display the spectrum of the observed temperature with foreground contamination removed, and the lower panels depict the power spectrum of the U-Net predicted field. Similar to the layout of Figure \ref{fig:multimK}, different columns correspond to results at various resolutions. As the box size of the field remain constant, finer resolutions leads to a larger range of Fourier modes. Due to the geometric shape of the foreground wedge, the observed maps at all resolutions lose a significant amount of information. 
Despite this, the U-Net model effectively restores power in the foreground-contaminated regions, closely approximating the original true signal across all scenarios presented. 
At lower resolutions, particularly for $64^3$, the amplitude of the restored power is noticeably lower than the original, especially near the edge of the foreground region. This discrepancy is most pronounced at large scales in the lower left corner of the plot. As resolution increases, this difference diminishes. Specifically, at comparable scales, such as $k\sim 0.3~{\rm h/Mpc}$, the restored power in higher resolution maps more closely approximates the true values than in lower resolution maps. However, a slight reduction in restored power remains observable diagonally near the boundary of the foreground wedge. 

\begin{figure}
    \centering
    \includegraphics[width=0.8\columnwidth]{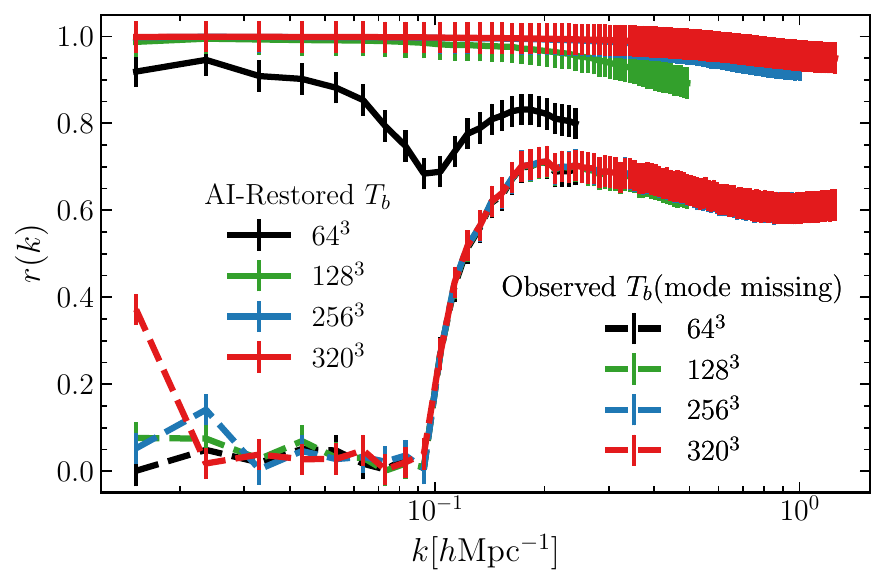}
    \caption{One-dimensional cross-correlation ratio between the true temperature field and the AI-restored fields (solid lines) as well as the observed $T_b$ fields (dashed lines) for maps with grid size of $64^3$, $128^3$, $256^3$, and $320^3$. Similar to Figure \ref{fig:multicof}, applying our AI model to fields with higher-resolution significantly enhance the cross-correlation ratio.  
    }
    \label{fig:ckop}
\end{figure}

Of course, the reemergence of power in the contaminated regions does not necessarily mean that the missing modes have been accurately restored. To assess this, we analyzed the correlation ratio between the AI-restored and original temperature distributions. The cross-correlation ratio between two fields $1$ and $2$ is defined as 
\begin{eqnarray}
     r(\vk) = \frac{P_{12}}{\sqrt{P_{1} P_{2}}} (\vk). 
\label{eq:corrat}
\end{eqnarray}
Here $P_{12} (\vk) = \langle \delta_{1} \delta_2 \rangle (\vk) $ represents the cross power spectrum between two fields, and $P_{1/2}(\vk)= \langle \delta_{1/2} \delta_{1/2} \rangle (\vk)$ is the respective auto power spectra. 
Figure \ref{fig:multicof} illustrates the cross-correlation ratio in two-dimensional $k_{\perp}-k_{\parallel}$ space, with resolutions varying from $64^3$ on the left to $320^3$ on the right. 
As shown, the cross-correlation ratio for the low-resolution $64^3$ sample is approximately $\sim 0.8$ at very large scales ($k\lesssim 0.1 ~{\rm h/Mpc}$) but drops to around $\sim 0.5$ at $k\sim 0.2-0.3 ~{\rm h/Mpc}$. Additionally, because the map is trained in the configuration space, the $r(\vk)$ for modes unaffected by the foreground in the upper left corner does not perfectly equal one. 
For the grid size of $128^3$, the correlation ratio with the true temperature map significantly improves, reaching around $0.8$ at $k \lesssim 0.3~{\rm h/Mpc}$. With larger grid sizes, the performance further improves. For example, the $256^3$ grid map achieves a correlation ratio of approximately $0.8$ at $k\sim 0.6 ~{\rm h/Mpc} $, \revone{compared to about $0.6 $ for the $128^3$ grid map}. This trend continues with a finer $320^3$ grid, which is the maximum size our GPU memory can accommodate. Therefore, it is reasonable to expect that, with more advanced GPU capabilities, the model could potentially achieve even better results.

Furthermore, we also present the one-dimensional correlation ratio $r(k)$, averaging over \rev{the} results in Figure \ref{fig:multicof}. As displayed in Figure \ref{fig:ckop}, solid lines represent the correlation ratio for the AI-restored fields, while dashed lines indicate the ratio for maps with missing modes due to foreground avoidance. Due to the extension of the foreground wedge to larger $k$ values, the 1D correlation ratio does not converge to one, even at smaller scales. Except for the lowest resolution map of $64^3$, the overall correlation ratio of the AI-restored fields reaches approximately $\sim 0.9$ at very small scales, $k\sim 1~{\rm h/Mpc}$. 
This performance exceeds the scale limits ($\sim 0.6 ~{\rm h/Mpc}$) where the BAO reconstruction algorithms typically cease to be effective, demonstrating that our AI restoration maintains sufficient fidelity for subsequent application of BAO reconstruction.

\subsection{BAO Reconstruction}
\label{sec:resbaorec}

\begin{figure}
    \includegraphics[width=1.0\columnwidth]{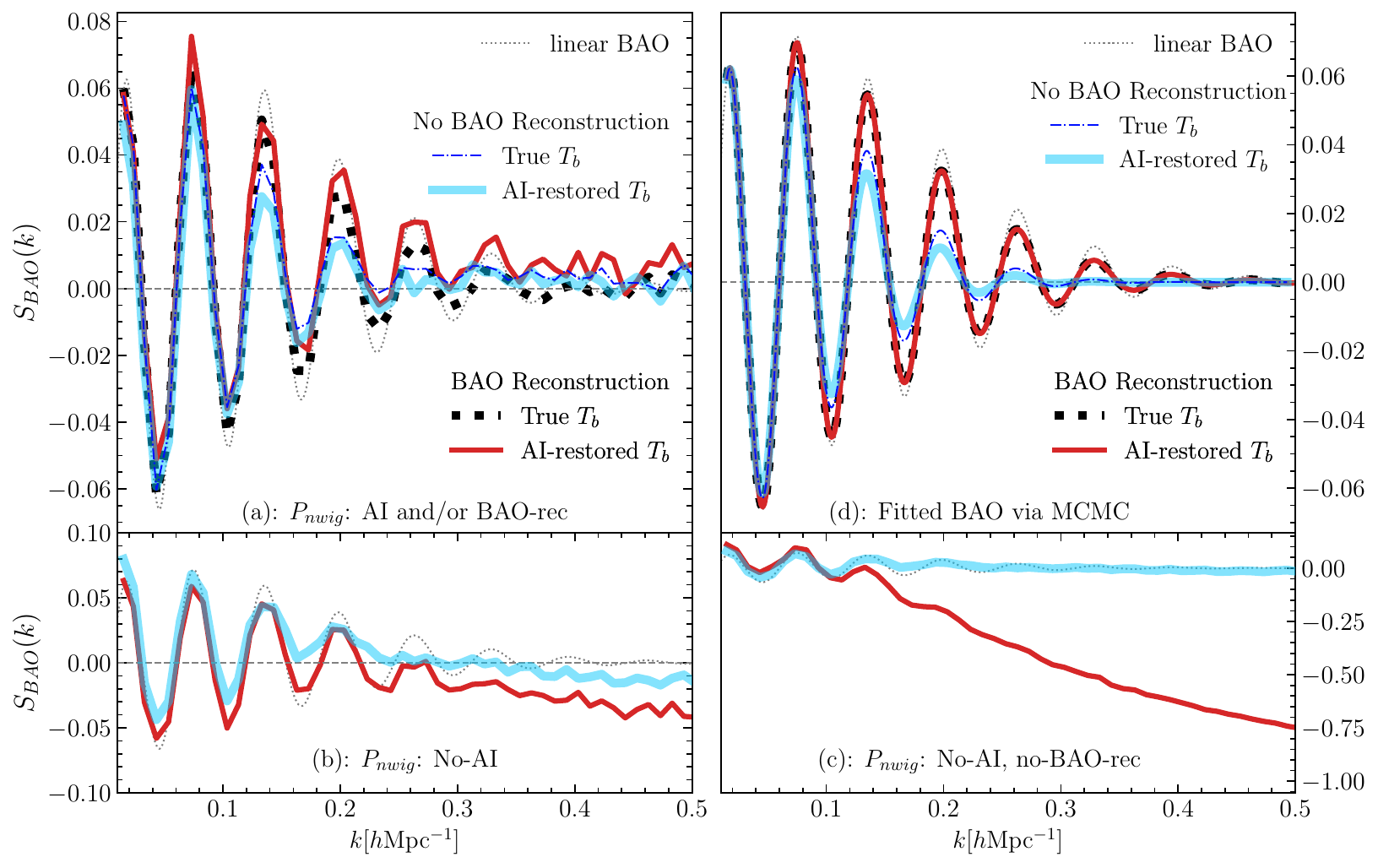}
    \caption{\revone{
    The BAO signatures of various fields. From panel (a) to (c), the BAO signature is obtained by dividing the power spectrum with BAO features by a seed-matched no-wiggle power spectrum, with each panel corresponding to a different level of processing applied to the no-wiggle spectrum.
    {\it Panel (a)}: The baseline result where the $P_\mathrm{nwig}$ has undergone the same AI restoration and/or BAO reconstruction process as the corresponding power spectrum with the BAO signature. As shown, in this most idealized case, AI restoration does not appear to significantly affect the performance of BAO reconstruction. 
    {\it Panel (b)}: A more realistic scenario where the no-wiggle spectrum has not undergone AI restoration. In this case, the AI restoration introduces mild shape deviations to the BAO signature. 
    {\it Panel (c)}: A scenario where the no-wiggle spectrum has undergone neither AI restoration nor BAO reconstruction. Now, the BAO reconstruction introduce quite significant shape deviation. 
    {\it Panel (d)}: BAO signals obtained from MCMC fitting of Equation (\ref{eqn:BAOsig}), with the smearing parameter $\Sigma$ determined from the fit. Here, the template $P_\mathrm{nwig}$ has not undergone any AI restoration or BAO reconstruction, and the BAO template $S_\mathrm{BAO}$ is defined using the theoretical linear model. }}
    \label{fig:bao_wiggles} 
\end{figure}

Having restored the lost Fourier modes, we proceed to reconstruct the BAO signature using our algorithm. 
\revone{As discussed in Section \ref{sec:paramfit}, we present the BAO result in two different ways.  First, as a baseline evaluation of the most idealized scenario}, we apply our trained AI model as well as the subsequent BAO reconstruction to a pair of simulations that share the same initial condition seed but differ in their transfer functions: one retains the BAO signature, while the other employs a `no-wiggle' smooth function.
\revone{The results are displayed in panel (a) of Figure \ref{fig:bao_wiggles}.} All signals, except for the linear model (thin black dotted line), are derived from data with a grid size of $256^3$. \rev{As seen, the BAO signal from the AI-restored field (thick sky-blue solid line) is very similar to the true 21cm brightness temperature map (thin blue dash-dotted line)}, indicating the effectiveness of our AI model. Due to the non-linear evolution, the wiggles after the third peaks are dampened away compared to the linear model. 
Similarly, after BAO reconstruction, we find that the AI-restored field (red solid line) closely aligns with the true post-reconstruction BAO signal that includes all modes (thick black dotted line). \revone{This suggests that, in this most idealized scenario, AI restoration does not appear to significantly impact the overall effectiveness of BAO reconstruction.} As shown, the BAO reconstruction enhances the BAO signature, particularly beyond the third peak, where it was previously dampened in the original field.

\revone{Of course, above result does not imply that the sharpened BAO peak can be extracted without difficulty. Since both the AI model and BAO reconstruction introduce scale-dependent distortions in the recovered power spectrum, additional care is required.
In a gradually more realistic scenario, shown in the second row of Figure \ref{fig:bao_wiggles}, we present the BAO signature obtained by dividing the processed power spectrum by a no-wiggle spectrum that has undergone no AI restoration (panel b), and one that has undergone neither AI restoration nor BAO reconstruction (panel c).
In panel (b), the sky-blue solid line is calculated by dividing the AI-restored power spectrum by a no-wiggle $P(k)$ without AI restoration. It shows that AI restoration alone introduces a mild shape distortion, leading to slight modifications in the BAO signature. These distortions are carried through to the BAO reconstruction, as seen in the red solid line, which is obtained by dividing the processed power spectrum by a no-wiggle spectrum that has undergone BAO reconstruction but not AI restoration. Nevertheless, it is evident that the BAO reconstruction still effectively reduces nonlinear smearing, allowing more BAO peaks to be recovered.
In panel (c), we further present the BAO signature obtained by dividing the processed power spectrum by a no-wiggle $P(k)$ that has undergone neither AI restoration nor BAO reconstruction. In this case, both lines are divided by the same unprocessed no-wiggle spectrum, which also serves as the no-wiggle template $P^\mathrm{temp}_\mathrm{nwig}(k)$ described in Equation (\ref{eqn:fittemp}). As a result, the sky-blue solid line in this panel is identical to the corresponding line in panel (b). 
However, the red solid line, representing the post-BAO-reconstruction result, shows that BAO reconstruction along introduces significantly greater shape deviations compared to AI restoration. Addressing these shape distortions introduced by BAO reconstruction is beyond the scope of this paper, but we plan to investigate them further and explore possible improvements in future work.
}


\revone{To isolate the BAO signature from the distorted curves above, we follow the fitting procedure described in Section \ref{sec:paramfit} and extract the smearing parameter $\Sigma$.
As previously mentioned, the template $P_\mathrm{nwig}^\mathrm{temp}(k)$ has not undergone any AI restoration or BAO reconstruction, and the BAO template $S^\mathrm{temp}_\mathrm{BAO}(k)$ is computed using the output of the Boltzmann code \texttt{CAMB} along with the corresponding fitted no-wiggle power spectrum. As a result, the fitted smearing parameter $\Sigma$ serves as a proxy for the amplitude of the BAO peaks, providing a quantitative measure of the (in)effectiveness of the BAO reconstruction.
}
Specifically, we employed the Markov Chain Monte Carlo (MCMC) method to fit the corresponding power spectrum. The extracted BAO signatures are presented in panel (d) of Figure \ref{fig:bao_wiggles}, with each curve showing the fitted BAO, as described by Equation (\ref{eqn:BAOsig}), and the corresponding $\Sigma$ value from the best fit to the simulated data. \revone{Compared to panel (a), this approach accurately captures the smearing parameter $\Sigma$, and the resulting BAO signal closely matches the corresponding result shown in panel (a).}
Additionally, in Appendix \ref{sec:appendix}, we also present corner plots of all parameters for two specific scenarios: the AI-restored $P(k)$ without BAO reconstruction (Figure \ref{fig:corner_pred_fit}) and with BAO reconstruction (Figure \ref{fig:corner_pred_rect_fit}). As shown, The posterior distribution of the polynomial coefficients $a_i$ is quite Gaussian, whereas the smearing coefficient exhibits some non-Gaussianity. Due to the large error on the power spectrum arising from the size of our simulation box, the one-$sigma$ error on $\Sigma$ is quite large. Nevertheless, as shown in Figure \ref{fig:bao_wiggles}, the best-fit BAO model remains sufficiently accurate.

One might notice that both reconstructed BAO signatures deviate from the linear theory (thin black dotted line), particularly at the second and third peaks. 
\rev{This deviation primarily arises from the imperfect de-biasing process (Equation \ref{eqn:debias}). Although the linear factor $\bar{T}_\mathrm{HI} b_\mathrm{T}$ has already been corrected for, the relationship between the brightness temperature fluctuation and the underlying dark matter is quite complex, various non-linear and non-local contributions may still reduce the effectiveness of the reconstruction.
} 
A secondary factor may be the resolution of the field. Constrained by the GPU limitation, the current grid size is not ideal for the BAO reconstruction. However, such effect is less significant compared to the clustering bias. Moreover, future advancements in computational capabilities are expected to mitigate the influence of resolution constraints on subsequent BAO reconstructions.

\section{Conclusions and Discussion}
\label{sec:conclusion}

With recent advancements in detecting the auto-power spectrum of 21cm intensity mapping \citep{Paul2023}, we are approaching closer to the ultimate goal of cosmological constraints using this technique. Despite these advances, many aspects of the data analysis process remain unexplored, potentially impacting future measurements. This paper focuses on the BAO reconstruction technique, originally developed for galaxy redshift surveys, which has evolved over many years to effectively reverse the effects of non-linear structure formation and restore the linear BAO signature, thereby enhancing the measurement accuracy of the sound horizon scale. However, foreground contamination severely compromises modes within the Fourier regions affected, including both intrinsic $k_{\parallel} < k_{\rm cut}$ and a wedge-like region, leading to the loss of valuable cosmological information and complicating the application of BAO reconstruction, which fundamentally operates in the configuration space. Fortunately, the non-linear structure formation, results in complex mode coupling, enabling the foreground-unaffected modes carry information from those that are missing.

In this paper, we utilize the U-Net deep learning architecture to retrieve this missing information and assess its impact on BAO reconstruction. We demonstrate that this popular and powerful deep learning architecture is capable of retrieving the missing information entangled within the observed data. Despite the success demonstrated here, there is potential for further improving the model by optimizing the loss function, introducing more effective \rev{hierarchical structures}, and providing more \rev{training datasets} etc.
During our training process, we discovered an interesting scale invariance when applying the AI model to large-scale structure data: a model trained on a coarser map could be successfully applied to finer resolutions, achieving even higher cross-correlation ratios. \rev{This indicates a successful restoration of the phase angles of the corresponding Fourier modes. One possible explanation is the scale invariance in mode coupling inherent in the nonlinear structure formation process. The deep learning model may be learning patterns at coarser grid resolutions that represent fundamental physical processes, allowing them to generalize across different scales.}  Furthermore, the architecture of U-Net, with its hierarchical structure of convolutional layers that capture features at multiple scales and then reconstruct the output at higher resolutions, is particularly well-suited to leveraging the scale-invariance of the large-scale structure data.
\rev{
However, this comes at the cost of introducing additional scale dependence in the power spectrum, as shown in Figure \ref{fig:multipk2d}. Further investigation is needed to fully understand the underlying cause of this effect.
}

\rev{However, as discussed earlier, our approach has several limitations, particularly in modeling the HI distribution for training samples. 
First, the mode coupling in COLA simulations may deviate from that in full N-body simulations, potentially affecting the AI model’s performance when applied to real data. Second, the low mass resolution of our dark matter field necessitates subgrid modeling of HI, which fails to preserve the spatial correlations present in a physically realistic HI distribution. This could introduce biases and alter mode coupling, particularly impacting AI restoration in high-resolution maps. Finally, our approximations in modeling redshift-space distortions may introduce additional deviations. In both resolved and unresolved halos, HI velocities deviate from those of their host galaxies or the underlying dark matter field, adding further uncertainties.
To improve our approach, future efforts could focus on enhancing the accuracy of both dark matter and HI field modeling. Additionally, given the challenges in generating large, high-precision training datasets, one could conduct consistency tests and apply the AI-trained model to hydrodynamical simulation data for further validation. 
}

\rev{
Furthermore, there are many systematic effects could potential affect the future application of this approach.} Crucially, for a machine learning-based method, it is essential to incorporate real-world observational effects into the training process to mitigate potential biases in the trained model. These include factors such as instrumental beam pattern, thermal noise, and residual foregrounds, among others.  We plan to address these aspects in future studies. 
Meanwhile, this is also the reason why we adopted this two-step approach of reconstruction, rather than directly predicting the linear fields from the foreground-contaminated temperature map. Given physical limitations set by the shell-crossing, the deep learning process is unlikely to significantly outperform traditional BAO reconstruction algorithms. Consequently, using AI in the subsequent process might introduce additional cosmology-dependent errors without significant benefits.

\acknowledgments
This work is supported by the National SKA Program of China (Grants Nos. 2022SKA0110200, 2022SKA0110202 and 2020SKA0110401), the National Science Foundation of China (Grants Nos. 12473006, 12373005, 12261141691), the China Manned Space Project with No. CMS-CSST-2021 (B01, A02, A03). The authors acknowledge the Beijing Super Cloud Center (BSCC, URL: http://www.blsc.cn/) for providing HPC resources that have contributed to the research results reported within this paper.

\appendix
\section{Parametric Fitting of BAO Signature}
\label{sec:appendix}
As discussed in Section \ref{sec:resbaorec}, we followed the standard approach for extracting the BAO damping parameter $\Sigma$ while accounting for the scale-dependent distortions introduced by the AI model. 
Specifically, Figure \ref{fig:corner_pred_fit} shows the multi-dimensional parameter distributions from the MCMC fitting of the AI-restored power spectrum without BAO reconstruction, while Figure \ref{fig:corner_pred_rect_fit} presents the corresponding results with both AI restoration and BAO reconstruction.

\begin{center}
    \includegraphics[width=0.55\columnwidth]{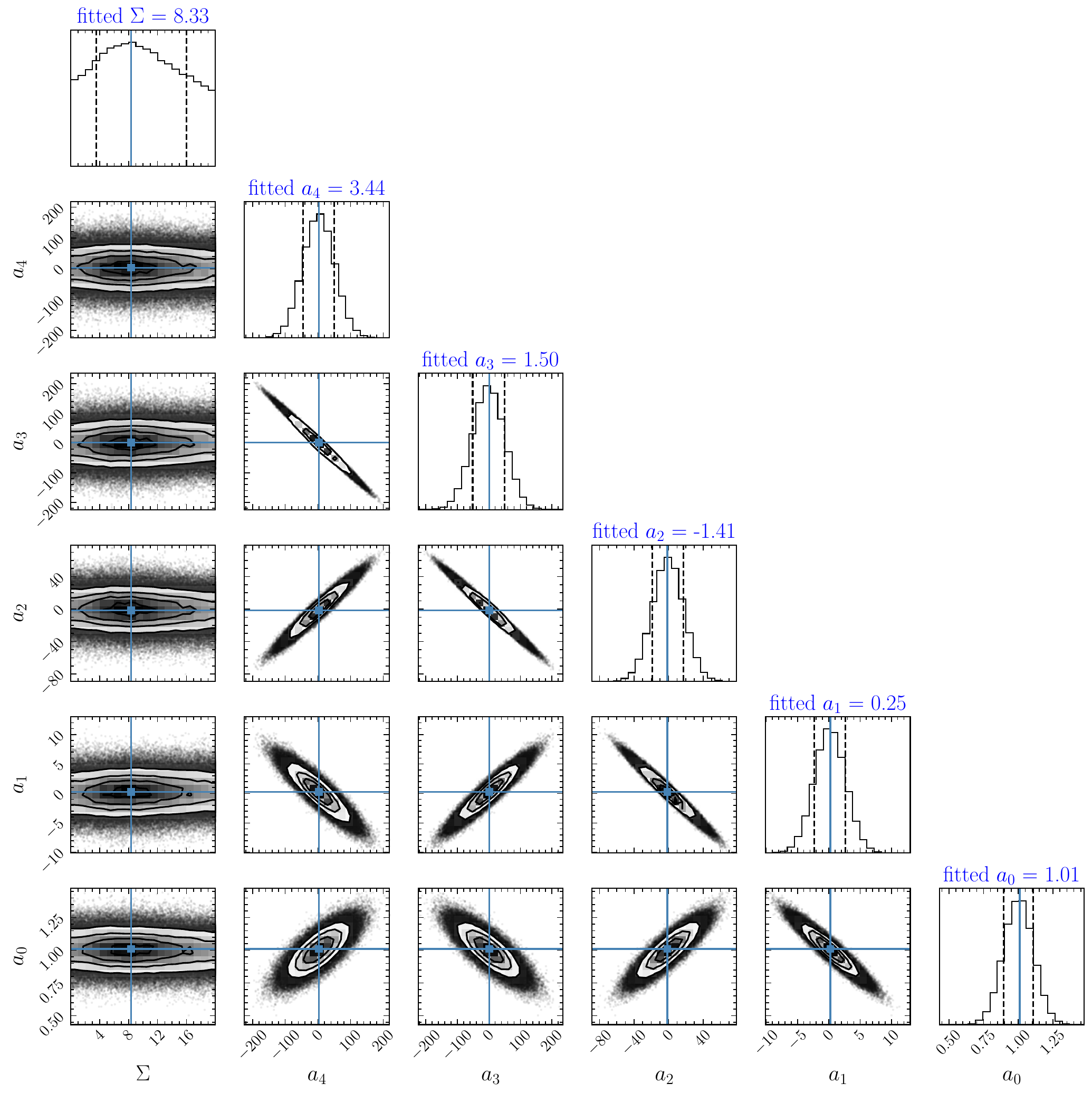}
    \captionof{figure}{\rev{The distribution of model parameters from the MCMC analysis for the AI-restored power spectrum without BAO reconstruction. The polynomial coefficients $a_i$ exhibit mostly Gaussian distributions, whereas the damping parameter $\Sigma$ shows some degree of non-Gaussianity. }}
    \label{fig:corner_pred_fit}
\end{center}

\begin{center}
    \includegraphics[width=0.55\columnwidth]{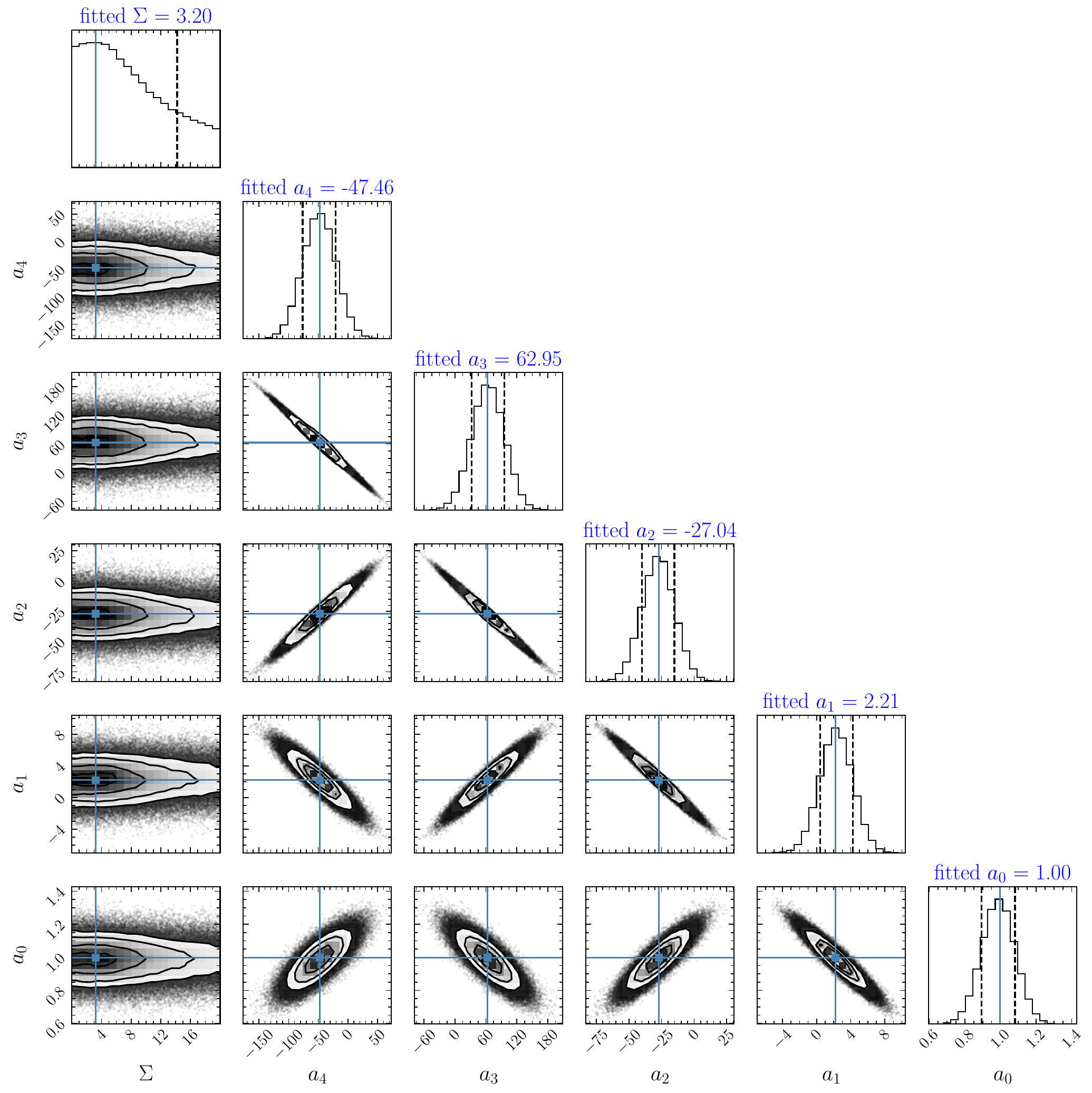}
    \captionof{figure}{\rev{Similar to Figure \ref{fig:corner_pred_fit}, the distribution of model parameters from the MCMC analysis for AI-restored power spectrum with BAO reconstruction.} }
    \label{fig:corner_pred_rect_fit}
\end{center}


\bibliographystyle{JHEP}
\bibliography{main.bib}

\providecommand{\href}[2]{#2}\begingroup\raggedright\begin{thebibliography}{10}

\bibitem{colless20012df}
M.~Colless, G.~Dalton, S.~Maddox, W.~Sutherland, P.~Norberg, S.~Cole et~al.,
  \emph{{The 2dF Galaxy Redshift Survey: spectra and redshifts}},
  \href{https://doi.org/10.1046/j.1365-8711.2001.04902.x}{\emph{Monthly Notices
  of the Royal Astronomical Society} {\bfseries 328} (2001) 1039–1063}.

\bibitem{jones20046df}
D.H.~Jones, W.~Saunders, M.~Colless, M.A.~Read, Q.A.~Parker, F.G.~Watson
  et~al., \emph{{The 6dF Galaxy Survey: samples, observational techniques and
  the first data release}},
  \href{https://doi.org/10.1111/j.1365-2966.2004.08353.x}{\emph{Monthly Notices
  of the Royal Astronomical Society} {\bfseries 355} (2004) 747–763}.

\bibitem{drinkwater2010wigglez}
M.J.~Drinkwater, R.J.~Jurek, C.~Blake, D.~Woods, K.A.~Pimbblet, K.~Glazebrook
  et~al., \emph{{The WiggleZ Dark Energy Survey: survey design and first data
  release}},
  \href{https://doi.org/10.1111/j.1365-2966.2009.15754.x}{\emph{Monthly Notices
  of the Royal Astronomical Society} {\bfseries 401} (2010) 1429–1452}.

\bibitem{anderson2014clustering}
L.~Anderson, {\'E}.~Aubourg, S.~Bailey, F.~Beutler, V.~Bhardwaj, M.~Blanton
  et~al., \emph{The clustering of galaxies in the {SDSS-III Baryon Oscillation
  Spectroscopic Survey: baryon acoustic oscillations in the Data Releases 10
  and 11 Galaxy samples}},
  \href{https://doi.org/10.1093/mnras/stu523}{\emph{Monthly Notices of the
  Royal Astronomical Society} {\bfseries 441} (2014) 24–62}.

\bibitem{dark2016dark}
D.E.S.~Collaboration:, T.~Abbott, F.~Abdalla, J.~Aleksi{\'c}, S.~Allam,
  A.~Amara et~al., \emph{{The Dark Energy Survey: more than dark energy--an
  overview}}, \href{https://doi.org/10.1093/mnras/stw641}{\emph{Monthly Notices
  of the Royal Astronomical Society} {\bfseries 460} (2016) 1270–1299}.

\bibitem{collaboration2023early}
D.~Collaboration, A.G.~Adame, J.~Aguilar, S.~Ahlen, S.~Alam, G.~Aldering
  et~al., \emph{The {Early} {Data} {Release} of the {Dark} {Energy}
  {Spectroscopic} {Instrument}},
  \href{https://doi.org/10.3847/1538-3881/ad3217}{\emph{The Astronomical
  Journal} {\bfseries 168} (2024) 58}.

\bibitem{ivezic2019lsst}
{\v{Z}}.~Ivezi\'{c}, S.M.~Kahn, J.A.~Tyson, B.~Abel, E.~Acosta, R.~Allsman
  et~al., \emph{{LSST}: {From} {Science} {Drivers} to {Reference} {Design} and
  {Anticipated} {Data} {Products}},
  \href{https://doi.org/10.3847/1538-4357/ab042c}{\emph{The Astrophysical
  Journal} {\bfseries 873} (2019) 111}.

\bibitem{bharadwaj_using_2001}
S.~Bharadwaj, B.B.~Nath and S.K.~Sethi, \emph{Using {HI} to probe large scale
  structures at z {\textasciitilde} 3},
  \href{https://doi.org/10.1007/BF02933588}{\emph{Journal of Astrophysics and
  Astronomy} {\bfseries 22} (2001) 21}.

\bibitem{battye_neutral_2004}
R.A.~Battye, R.D.~Davies and J.~Weller, \emph{Neutral hydrogen surveys for high
  redshift galaxy clusters and proto-clusters},
  \href{https://doi.org/10.1111/j.1365-2966.2004.08416.x}{\emph{Monthly Notices
  of the Royal Astronomical Society} {\bfseries 355} (2004) 1339}.

\bibitem{furlanetto_cosmology_2006}
S.~Furlanetto, S.P.~Oh and F.~Briggs, \emph{Cosmology at {Low} {Frequencies}:
  {The} 21 cm {Transition} and the {High}-{Redshift} {Universe}},
  \href{https://doi.org/10.1016/j.physrep.2006.08.002}{\emph{Physics Reports}
  {\bfseries 433} (2006) 181}.

\bibitem{chang_intensity_2010}
T.-C.~Chang, U.-L.~Pen, K.~Bandura and J.B.~Peterson, \emph{An intensity map of
  hydrogen 21-cm emission at redshift z $\approx$ 0.8},
  \href{https://doi.org/10.1038/nature09187}{\emph{Nature} {\bfseries 466}
  (2010) 463}.

\bibitem{morales_reionization_2010}
M.F.~Morales and J.S.B.~Wyithe, \emph{Reionization and {Cosmology} with 21 cm
  {Fluctuations}},
  \href{https://doi.org/10.1146/annurev-astro-081309-130936}{\emph{Annual
  Review of Astronomy and Astrophysics} {\bfseries 48} (2010) }.

\bibitem{masui_measurement_2013}
K.W.~Masui, E.R.~Switzer, N.~Banavar, K.~Bandura, C.~Blake, L.-M.~Calin et~al.,
  \emph{{MEASUREMENT} {OF} 21 cm {BRIGHTNESS} {FLUCTUATIONS} {AT} \textit{z}
  $\sim$ 0.8 {IN} {CROSS}-{CORRELATION}},
  \href{https://doi.org/10.1088/2041-8205/763/1/L20}{\emph{The Astrophysical
  Journal} {\bfseries 763} (2013) L20}.

\bibitem{bull_late-time_2015}
P.~Bull, P.G.~Ferreira, P.~Patel and M.G.~Santos, \emph{Late-time cosmology
  with 21cm intensity mapping experiments},
  \href{https://doi.org/10.1088/0004-637X/803/1/21}{\emph{The Astrophysical
  Journal} {\bfseries 803} (2015) 21}.

\bibitem{santos_cosmology_2015}
M.~Santos, P.~Bull, D.~Alonso, S.~Camera, P.~Ferreira, G.~Bernardi et~al.,
  \emph{Cosmology with a {SKA} {HI} intensity mapping survey},
  {\emph{Proceedings of Science} {\bfseries 9-13-June-2014} (2014) }.

\bibitem{villaescusa-navarro_ingredients_2018}
F.~Villaescusa-Navarro, S.~Genel, E.~Castorina, A.~Obuljen, D.N.~Spergel,
  L.~Hernquist et~al., \emph{Ingredients for 21cm intensity mapping},
  \href{https://doi.org/10.3847/1538-4357/aadba0}{\emph{The Astrophysical
  Journal} {\bfseries 866} (2018) 135}.

\bibitem{lidz_intensity_2011}
A.~Lidz, S.R.~Furlanetto, S.P.~Oh, J.~Aguirre, T.-C.~Chang, O.~Doré et~al.,
  \emph{Intensity {Mapping} with {Carbon} {Monoxide} {Emission} {Lines} and the
  {Redshifted} 21 cm {Line}},
  \href{https://doi.org/10.1088/0004-637X/741/2/70}{\emph{The Astrophysical
  Journal} {\bfseries 741} (2011) 70}.

\bibitem{camera_cosmology_2013}
S.~Camera, M.G.~Santos, P.G.~Ferreira and L.~Ferramacho, \emph{Cosmology on
  {Ultralarge} {Scales} with {Intensity} {Mapping} of the {Neutral} {Hydrogen}
  21 cm {Emission}: {Limits} on {Primordial} {Non}-{Gaussianity}},
  \href{https://doi.org/10.1103/PhysRevLett.111.171302}{\emph{Physical Review
  Letters} {\bfseries 111} (2013) 171302}.

\bibitem{group_cosmology_2020}
S.K.A.C.S.W.~Group, D.J.~Bacon, R.A.~Battye, P.~Bull, S.~Camera, P.G.~Ferreira
  et~al., \emph{Cosmology with {Phase} 1 of the {Square} {Kilometre} {Array};
  {Red} {Book} 2018: {Technical} specifications and performance forecasts},
  \href{https://doi.org/10.1017/pasa.2019.51}{\emph{Publications of the
  Astronomical Society of Australia} {\bfseries 37} (2020) e007}.

\bibitem{wang_hi_2021}
J.~Wang, M.G.~Santos, P.~Bull, K.~Grainge, S.~Cunnington, J.~Fonseca et~al.,
  \emph{{HI} intensity mapping with {MeerKAT}: {Calibration} pipeline for
  multi-dish autocorrelation observations},
  \href{https://doi.org/10.1093/mnras/stab1365}{\emph{Monthly Notices of the
  Royal Astronomical Society} {\bfseries 505} (2021) 3698}.

\bibitem{schlegel_spectroscopic_2022}
D.J.~Schlegel, S.~Ferraro, G.~Aldering, C.~Baltay, S.~BenZvi, R.~Besuner
  et~al., \emph{A {Spectroscopic Road Map for Cosmic Frontier}: {DESI},
  {DESI-II}, {Stage-5}}, {\emph{arXiv e-prints} (2022) }
  [\href{https://arxiv.org/abs/arXiv:2209.03585}{{\ttfamily
  arXiv:2209.03585}}].

\bibitem{collaboration_euclid_2024}
E.~Collaboration, Y.~Mellier, Abdurro'uf, J.A.A.~Barroso, A.~Achúcarro,
  J.~Adamek et~al., \emph{Euclid. {I}. {Overview} of the {Euclid} mission},
  {\emph{arXiv e-prints} (2024) }
  [\href{https://arxiv.org/abs/arXiv:2405.13491}{{\ttfamily
  arXiv:2405.13491}}].

\bibitem{liu_data_2020}
A.~Liu and J.R.~Shaw, \emph{Data {Analysis} for {Precision} 21 cm {Cosmology}},
  \href{https://doi.org/10.1088/1538-3873/ab5bfd}{\emph{Publications of the
  Astronomical Society of the Pacific} {\bfseries 132} (2020) 062001}.

\bibitem{GSM2008MNRAS}
A.~{de Oliveira-Costa}, M.~{Tegmark}, B.M.~{Gaensler}, J.~{Jonas},
  T.L.~{Landecker} and P.~{Reich}, \emph{{A model of diffuse Galactic radio
  emission from 10 MHz to 100 GHz}},
  \href{https://doi.org/10.1111/j.1365-2966.2008.13376.x}{\emph{Monthly Notices
  of the Royal Astronomical Society} {\bfseries 388} (2008) 247}
  [\href{https://arxiv.org/abs/arXiv:0802.1525}{{\ttfamily arXiv:0802.1525}}].

\bibitem{GSM2017MNRAS}
H.~{Zheng}, M.~{Tegmark}, J.S.~{Dillon}, D.A.~{Kim}, A.~{Liu}, A.R.~{Neben}
  et~al., \emph{{{An improved model of diffuse galactic radio emission from 10
  MHz to 5 THz}}}, \href{https://doi.org/10.1093/mnras/stw2525}{\emph{Monthly
  Notices of the Royal Astronomical Society} {\bfseries 464} (2017) 3486}
  [\href{https://arxiv.org/abs/arXiv:1605.04920}{{\ttfamily
  arXiv:1605.04920}}].

\bibitem{Ekers_Bell_2002}
R.~Ekers and J.~Bell, \emph{{Radio Frequency Interference}},
  \href{https://doi.org/10.1017/S0074180900169669}{\emph{Symposium -
  International Astronomical Union} {\bfseries 199} (2002) 498–505}.

\bibitem{Ding2024ApJS}
J.~{Ding}, X.~{Wang}, U.-L.~{Pen} and X.-D.~{Li}, \emph{{Correlation-based Beam
  Calibration of 21 cm Intensity Mapping}},
  \href{https://doi.org/10.3847/1538-4365/ad6f0a}{\emph{The Astrophysical
  Journal Supplement Series} {\bfseries 274} (2024) 44}
  [\href{https://arxiv.org/abs/arXiv:2408.06682}{{\ttfamily
  arXiv:2408.06682}}].

\bibitem{Santos2005ApJ}
M.G.~{Santos}, A.~{Cooray} and L.~{Knox}, \emph{{Multifrequency Analysis of 21
  Centimeter Fluctuations from the Era of Reionization}},
  \href{https://doi.org/10.1086/429857}{\emph{The Astrophysical Journal}
  {\bfseries 625} (2005) 575}
  [\href{https://arxiv.org/abs/astro-ph/0408515}{{\ttfamily
  astro-ph/0408515}}].

\bibitem{Liu2009MNRAS}
A.~{Liu}, M.~{Tegmark}, J.~{Bowman}, J.~{Hewitt} and M.~{Zaldarriaga},
  \emph{{An improved method for 21-cm foreground removal}},
  \href{https://doi.org/10.1111/j.1365-2966.2009.15156.x}{\emph{Monthly Notices
  of the Royal Astronomical Society} {\bfseries 398} (2009) 401}
  [\href{https://arxiv.org/abs/arXiv:0903.4890}{{\ttfamily arXiv:0903.4890}}].

\bibitem{switzer2013determination}
E.R.~Switzer, K.W.~Masui, K.~Bandura, L.-M.~Calin, T.-C.~Chang, X.-L.~Chen
  et~al., \emph{Determination of z $\sim$ 0.8 neutral hydrogen fluctuations
  using the 21cm intensity mapping autocorrelation},
  \href{https://doi.org/10.1093/mnrasl/slt074}{\emph{Monthly Notices of the
  Royal Astronomical Society: Letters} {\bfseries 434} (2013) L46–L50}.

\bibitem{switzer_interpreting_2015}
E.R.~Switzer, T.-C.~Chang, K.W.~Masui, U.-L.~Pen and T.C.~Voytek, \emph{{
  INTERPRETING THE UNRESOLVED INTENSITY OF COSMOLOGICALLY REDSHIFTED LINE
  RADIATION}}, \href{https://doi.org/10.1088/0004-637X/815/1/51}{\emph{The
  Astrophysical Journal} {\bfseries 815} (2015) 51}.

\bibitem{bigot2015simulations}
M.-A.~Bigot-Sazy, C.~Dickinson, R.A.~Battye, I.W.A.~Browne, Y.-Z.~Ma, B.~Maffei
  et~al., \emph{Simulations for single-dish intensity mapping experiments},
  \href{https://doi.org/10.1093/mnras/stv2153}{\emph{Monthly Notices of the
  Royal Astronomical Society} {\bfseries 454} (2015) 3240}.

\bibitem{chapman2012foreground}
E.~Chapman, F.B.~Abdalla, G.~Harker, V.~Jelić, P.~Labropoulos, S.~Zaroubi
  et~al., \emph{{Foreground Removal using FastICA: A Showcase of LOFAR-EoR}},
  \href{https://doi.org/10.1111/j.1365-2966.2012.21065.x}{\emph{Monthly Notices
  of the Royal Astronomical Society} {\bfseries 423} (2012) 2518–2532}.

\bibitem{wolz2014effect}
L.~Wolz, F.B.~Abdalla, C.~Blake, J.R.~Shaw, E.~Chapman and S.~Rawlings,
  \emph{{The effect of foreground subtraction on cosmological measurements from
  intensity mapping}},
  \href{https://doi.org/10.1093/mnras/stu792}{\emph{Monthly Notices of the
  Royal Astronomical Society} {\bfseries 441} (2014) 3271–3283}.

\bibitem{eisenstein_improving_2007}
D.J.~Eisenstein, H.~Seo, E.~Sirko and D.N.~Spergel, \emph{Improving
  {Cosmological} {Distance} {Measurements} by {Reconstruction} of the {Baryon}
  {Acoustic} {Peak}}, \href{https://doi.org/10.1086/518712}{\emph{The
  Astrophysical Journal} {\bfseries 664} (2007) 675}.

\bibitem{ZYP17a}
H.-M.~{Zhu}, Y.~{Yu}, U.-L.~{Pen}, X.~{Chen} and H.-R.~{Yu}, \emph{{Nonlinear
  reconstruction}},
  \href{https://doi.org/10.1103/PhysRevD.96.123502}{\emph{Physical Review D}
  {\bfseries 96} (2017) 123502}
  [\href{https://arxiv.org/abs/arXiv:1611.09638}{{\ttfamily
  arXiv:1611.09638}}].

\bibitem{Yu2017a}
Y.~{Yu}, H.-M.~{Zhu} and U.-L.~{Pen}, \emph{{Halo Nonlinear Reconstruction}},
  \href{https://doi.org/10.3847/1538-4357/aa89e7}{\emph{The Astrophysical
  Journal} {\bfseries 847} (2017) 110}
  [\href{https://arxiv.org/abs/arXiv:1703.08301}{{\ttfamily
  arXiv:1703.08301}}].

\bibitem{Wang17}
X.~{Wang}, H.-R.~{Yu}, H.-M.~{Zhu}, Y.~{Yu}, Q.~{Pan} and U.-L.~{Pen},
  \emph{{Isobaric Reconstruction of the Baryonic Acoustic Oscillation}},
  \href{https://doi.org/10.3847/2041-8213/aa738c}{\emph{The Astrophysical
  Journall} {\bfseries 841} (2017) L29}
  [\href{https://arxiv.org/abs/arXiv:1703.09742}{{\ttfamily
  arXiv:1703.09742}}].

\bibitem{SBZ17}
M.~{Schmittfull}, T.~{Baldauf} and M.~{Zaldarriaga}, \emph{{Iterative initial
  condition reconstruction}},
  \href{https://doi.org/10.1103/PhysRevD.96.023505}{\emph{Physical Review D}
  {\bfseries 96} (2017) 023505}
  [\href{https://arxiv.org/abs/arXiv:1704.06634}{{\ttfamily
  arXiv:1704.06634}}].

\bibitem{obuljen_baryon_2017}
A.~Obuljen, F.~Villaescusa-Navarro, E.~Castorina and M.~Viel, \emph{Baryon
  {Acoustic} {Oscillations} reconstruction with pixels},
  \href{https://doi.org/10.1088/1475-7516/2017/09/012}{\emph{Journal of
  Cosmology and Astroparticle Physics} {\bfseries 2017} (2017) 012}.

\bibitem{Wang_2019}
X.~Wang and U.-L.~Pen, \emph{{Understanding the Reconstruction of the Biased
  Tracer}}, \href{https://doi.org/10.3847/1538-4357/aaf231}{\emph{The
  Astrophysical Journal} {\bfseries 870} (2019) 116}.

\bibitem{BJLi2019MNRAS}
J.~{Birkin}, B.~{Li}, M.~{Cautun} and Y.~{Shi}, \emph{{Reconstructing the
  baryon acoustic oscillations using biased tracers}},
  \href{https://doi.org/10.1093/mnras/sty3365}{\emph{Monthly Notices of the
  Royal Astronomical Society} {\bfseries 483} (2019) 5267}
  [\href{https://arxiv.org/abs/arXiv:1809.08135}{{\ttfamily
  arXiv:1809.08135}}].

\bibitem{seo_foreground_2016}
H.-J.~Seo and C.M.~Hirata, \emph{The foreground wedge and 21 cm {BAO} surveys},
  \href{https://doi.org/10.1093/mnras/stv2806}{\emph{Monthly Notices of the
  Royal Astronomical Society} {\bfseries 456} (2016) 3142}.

\bibitem{cohn_combining_2016}
J.D.~Cohn, M.~White, T.-C.~Chang, G.~Holder, N.~Padmanabhan and O.~Doré,
  \emph{Combining galaxy and 21cm surveys},
  \href{https://doi.org/10.1093/mnras/stw108}{\emph{Monthly Notices of the
  Royal Astronomical Society} {\bfseries 457} (2016) 2068}.

\bibitem{zhu2016}
H.-M.~Zhu, U.-L.~Pen, Y.~Yu, X.~Er and X.~Chen, \emph{{Cosmic Tidal
  Reconstruction}},
  \href{https://doi.org/10.1103/PhysRevD.93.103504}{\emph{Physical Review D}
  {\bfseries 93} (2016) 103504}.

\bibitem{villaescusa-navarro_weighing_2015}
F.~Villaescusa-Navarro, P.~Bull and M.~Viel, \emph{Weighing neutrinos with
  cosmic neutral hydrogen},
  \href{https://doi.org/10.1088/0004-637X/814/2/146}{\emph{The Astrophysical
  Journal} {\bfseries 814} (2015) 146}.

\bibitem{sadeh_annz2_2016}
I.~Sadeh, F.B.~Abdalla and O.~Lahav, \emph{{ANNz2}: {Photometric} {Redshift}
  and {Probability} {Distribution} {Function} {Estimation} using {Machine}
  {Learning}},
  \href{https://doi.org/10.1088/1538-3873/128/968/104502}{\emph{Publications of
  the Astronomical Society of the Pacific} {\bfseries 128} (2016) 104502}.

\bibitem{ho_robust_2019}
M.~Ho, M.M.~Rau, M.~Ntampaka, A.~Farahi, H.~Trac and B.~Poczos, \emph{A
  {Robust} and {Efficient} {Deep} {Learning} {Method} for {Dynamical} {Mass}
  {Measurements} of {Galaxy} {Clusters}},
  \href{https://doi.org/10.3847/1538-4357/ab4f82}{\emph{The Astrophysical
  Journal} {\bfseries 887} (2019) 25}.

\bibitem{agarap_deep_2019}
A.F.~Agarap, \emph{Deep learning using rectified linear units (relu)},
  {\emph{arXiv e-prints} (2018) }
  [\href{https://arxiv.org/abs/arXiv:1803.08375}{{\ttfamily
  arXiv:1803.08375}}].

\bibitem{wu_cosmic_2021}
Z.~Wu, Z.~Zhang, S.~Pan, H.~Miao, X.~Luo, X.~Wang et~al., \emph{Cosmic
  {Velocity} {Field} {Reconstruction} {Using} {AI}},
  \href{https://doi.org/10.3847/1538-4357/abf3bb}{\emph{The Astrophysical
  Journal} {\bfseries 913} (2021) 2}.

\bibitem{villanueva-domingo_inferring_2022}
P.~Villanueva-Domingo, F.~Villaescusa-Navarro, D.~Anglés-Alcázar, S.~Genel,
  F.~Marinacci, D.N.~Spergel et~al., \emph{Inferring {Halo} {Masses} with
  {Graph} {Neural} {Networks}},
  \href{https://doi.org/10.3847/1538-4357/ac7aa3}{\emph{The Astrophysical
  Journal} {\bfseries 935} (2022) 30}.

\bibitem{schaurecker_super-resolving_2022}
D.~Schaurecker, Y.~Li, J.~Tinker, S.~Ho and A.~Refregier, \emph{Super-resolving
  {Dark} {Matter} {Halos} using {Generative} {Deep} {Learning}}, {\emph{arXiv
  e-prints} (2022) } [\href{https://arxiv.org/abs/arXiv:2111.06393}{{\ttfamily
  arXiv:2111.06393}}].

\bibitem{shi_21_2024}
F.~Shi, H.~Chang, L.~Zhang, H.~Shan, J.~Zhang, S.~Zhou et~al., \emph{21-cm
  foreground removal using ai and the frequency-difference technique},
  \href{https://doi.org/10.1103/physrevd.109.063509}{\emph{Physical Review D}
  {\bfseries 109} (2024) }.

\bibitem{masipa_emulating_2023}
M.P.~Masipa, S.~Hassan, M.G.~Santos, G.~Contardo and K.~Cho, \emph{Emulating
  radiation transport on cosmological scale using a denoising {Unet}},
  {\emph{arXiv e-prints} (2023) }
  [\href{https://arxiv.org/abs/arXiv:2303.12065}{{\ttfamily
  arXiv:2303.12065}}].

\bibitem{mao_baryon_2020}
T.-X.~Mao, J.~Wang, B.~Li, Y.-C.~Cai, B.~Falck, M.~Neyrinck et~al.,
  \emph{Baryon acoustic oscillations reconstruction using convolutional neural
  networks}, \href{https://doi.org/10.1093/mnras/staa3741}{\emph{Monthly
  Notices of the Royal Astronomical Society} {\bfseries 501} (2020) 1499}.

\bibitem{li_separating_2019}
W.~Li, H.~Xu, Z.~Ma, R.~Zhu, D.~Hu, Z.~Zhu et~al., \emph{Separating the {EoR}
  signal with a convolutional denoising autoencoder: a deep-learning-based
  method}, \href{https://doi.org/10.1093/mnras/stz582}{\emph{Monthly Notices of
  the Royal Astronomical Society} {\bfseries 485} (2019) 2628}.

\bibitem{gagnon-hartman_recovering_2021}
S.~Gagnon-Hartman, Y.~Cui, J.~Kennedy, A.~Liu and S.~Ravanbakhsh,
  \emph{Recovering the {Wedge} {Modes} {Lost} to 21-cm {Foregrounds}},
  \href{https://doi.org/10.1093/mnras/stab1158}{\emph{Monthly Notices of the
  Royal Astronomical Society} {\bfseries 504} (2021) 4716}.

\bibitem{bianco_deep_2021}
M.~Bianco, S.K.~Giri, I.T.~Iliev and G.~Mellema, \emph{Deep learning approach
  for identification of {HII} regions during reionization in 21-cm
  observations}, \href{https://doi.org/10.1093/mnras/stab1518}{\emph{Monthly
  Notices of the Royal Astronomical Society} {\bfseries 505} (2021) 3982}.

\bibitem{datta_bright_2010}
A.~Datta, J.D.~Bowman and C.L.~Carilli, \emph{{BRIGHT} {SOURCE} {SUBTRACTION}
  {REQUIREMENTS} {FOR} {REDSHIFTED} 21 cm {MEASUREMENTS}},
  \href{https://doi.org/10.1088/0004-637X/724/1/526}{\emph{The Astrophysical
  Journal} {\bfseries 724} (2010) 526}.

\bibitem{morales_four_2012}
M.F.~Morales, B.~Hazelton, I.~Sullivan and A.~Beardsley, \emph{{FOUR}
  {FUNDAMENTAL} {FOREGROUND} {POWER} {SPECTRUM} {SHAPES} {FOR} 21 cm
  {COSMOLOGY} {OBSERVATIONS}},
  \href{https://doi.org/10.1088/0004-637X/752/2/137}{\emph{The Astrophysical
  Journal} {\bfseries 752} (2012) 137}.

\bibitem{parsons_per-baseline_2012}
A.R.~Parsons, J.C.~Pober, J.E.~Aguirre, C.L.~Carilli, D.C.~Jacobs and
  D.F.~Moore, \emph{A {Per}-{Baseline}, {Delay}-{Spectrum} {Technique} for
  {Accessing} the 21cm {Cosmic} {Reionization} {Signature}},
  \href{https://doi.org/10.1088/0004-637X/756/2/165}{\emph{The Astrophysical
  Journal} {\bfseries 756} (2012) 165}.

\bibitem{BAO2007ApJEis}
D.J.~{Eisenstein}, H.-J.~{Seo} and M.~{White}, \emph{{On the Robustness of the
  Acoustic Scale in the Low-Redshift Clustering of Matter}},
  \href{https://doi.org/10.1086/518755}{\emph{The Astrophysical Journal}
  {\bfseries 664} (2007) 660}
  [\href{https://arxiv.org/abs/astro-ph/0604361}{{\ttfamily
  astro-ph/0604361}}].

\bibitem{SDSSrec2012MNRAS}
K.T.~{Mehta}, A.J.~{Cuesta}, X.~{Xu}, D.J.~{Eisenstein} and N.~{Padmanabhan},
  \emph{{A 2 per cent distance to z = 0.35 by reconstructing baryon acoustic
  oscillations - III. Cosmological measurements and interpretation}},
  \href{https://doi.org/10.1111/j.1365-2966.2012.21112.x}{\emph{Monthly Notices
  of the Royal Astronomical Society} {\bfseries 427} (2012) 2168}
  [\href{https://arxiv.org/abs/arXiv:1202.0092}{{\ttfamily arXiv:1202.0092}}].

\bibitem{SDSS3rec2016MNRAS}
H.~{Gil-Mar{\'\i}n}, W.J.~{Percival}, A.J.~{Cuesta}, J.R.~{Brownstein},
  C.-H.~{Chuang}, S.~{Ho} et~al., \emph{{The clustering of galaxies in the
  SDSS-III Baryon Oscillation Spectroscopic Survey: BAO measurement from the
  LOS-dependent power spectrum of DR12 BOSS galaxies}},
  \href{https://doi.org/10.1093/mnras/stw1264}{\emph{Monthly Notices of the
  Royal Astronomical Society} {\bfseries 460} (2016) 4210}
  [\href{https://arxiv.org/abs/arXiv:1509.06373}{{\ttfamily
  arXiv:1509.06373}}].

\bibitem{SDSS4rec2018ApJ}
J.E.~{Bautista}, M.~{Vargas-Maga{\~n}a}, K.S.~{Dawson}, W.J.~{Percival},
  J.~{Brinkmann}, J.~{Brownstein} et~al., \emph{{The SDSS-IV Extended Baryon
  Oscillation Spectroscopic Survey: Baryon Acoustic Oscillations at Redshift of
  0.72 with the DR14 Luminous Red Galaxy Sample}},
  \href{https://doi.org/10.3847/1538-4357/aacea5}{\emph{The Astrophysical
  Journal} {\bfseries 863} (2018) 110}
  [\href{https://arxiv.org/abs/arXiv:1712.08064}{{\ttfamily
  arXiv:1712.08064}}].

\bibitem{DESIBAO2024arXiv}
{DESI Collaboration}, A.G.~{Adame}, J.~{Aguilar}, S.~{Ahlen}, S.~{Alam},
  D.M.~{Alexander} et~al., \emph{{DESI 2024 III: Baryon Acoustic Oscillations
  from Galaxies and Quasars}},
  \href{https://doi.org/10.48550/arXiv.2404.03000}{\emph{arXiv e-prints} (2024)
  } [\href{https://arxiv.org/abs/arXiv:2404.03000}{{\ttfamily
  arXiv:2404.03000}}].

\bibitem{COLA2013JCAP}
S.~{Tassev}, M.~{Zaldarriaga} and D.J.~{Eisenstein}, \emph{{Solving large scale
  structure in ten easy steps with COLA}},
  \href{https://doi.org/10.1088/1475-7516/2013/06/036}{\emph{JCAP} {\bfseries
  2013} (2013) 036} [\href{https://arxiv.org/abs/arXiv:1301.0322}{{\ttfamily
  arXiv:1301.0322}}].

\bibitem{behroozi_rockstar_2013}
P.S.~Behroozi, R.H.~Wechsler and H.-Y.~Wu, \emph{The {Rockstar} {Phase}-{Space}
  {Temporal} {Halo} {Finder} and the {Velocity} {Offsets} of {Cluster}
  {Cores}}, \href{https://doi.org/10.1088/0004-637X/762/2/109}{\emph{The
  Astrophysical Journal} {\bfseries 762} (2013) 109}.

\bibitem{CoorySheth2002}
A.~Cooray and R.~Sheth, \emph{{Halo models of large scale structure}},
  \href{https://doi.org/10.1016/S0370-1573(02)00276-4}{\emph{Physics Reports}
  {\bfseries 372} (2002) 1}.

\bibitem{ronneberger_u-net_2015}
O.~Ronneberger, P.~Fischer and T.~Brox, \emph{U-{Net}: {Convolutional}
  {Networks} for {Biomedical} {Image} {Segmentation}},  in \emph{{MICCAI}
  2015}, (Cham), pp.~234--241, Springer International Publishing, 2015
  [\href{https://arxiv.org/abs/arXiv:1505.04597}{{\ttfamily
  arXiv:1505.04597}}].

\bibitem{shelhamer_fully_2017}
E.~Shelhamer, J.~Long and T.~Darrell, \emph{{Fully Convolutional Networks for
  Semantic Segmentation}},
  \href{https://doi.org/10.1109/TPAMI.2016.2572683}{\emph{IEEE Transactions on
  Pattern Analysis and Machine Intelligence} {\bfseries 39} (2017) 640}.

\bibitem{kingma_adam_2017}
D.P.~Kingma and J.~Ba, \emph{{Adam: A Method for Stochastic Optimization}},
  [\href{https://arxiv.org/abs/arXiv:1412.6980}{{\ttfamily arXiv:1412.6980}}].

\bibitem{Schmittfull_2017PhRvD}
M.~{Schmittfull}, T.~{Baldauf} and M.~{Zaldarriaga}, \emph{{Iterative initial
  condition reconstruction}},
  \href{https://doi.org/10.1103/PhysRevD.96.023505}{\emph{Physical Review D}
  {\bfseries 96} (2017) 023505}
  [\href{https://arxiv.org/abs/arXiv:1704.06634}{{\ttfamily
  arXiv:1704.06634}}].

\bibitem{SDSS3BAO_2012MNRAS}
L.~{Anderson}, E.~{Aubourg}, S.~{Bailey}, D.~{Bizyaev}, M.~{Blanton},
  A.S.~{Bolton} et~al., \emph{{The clustering of galaxies in the SDSS-III
  Baryon Oscillation Spectroscopic Survey: baryon acoustic oscillations in the
  Data Release 9 spectroscopic galaxy sample}},
  \href{https://doi.org/10.1111/j.1365-2966.2012.22066.x}{\emph{MNRAS}
  {\bfseries 427} (2012) 3435}
  [\href{https://arxiv.org/abs/arXiv1203.6594}{{\ttfamily arXiv1203.6594}}].

\bibitem{2011ascl.soft02026L}
A.~{Lewis} and A.~{Challinor}, ``{CAMB: Code for Anisotropies in the Microwave
  Background}.'' Astrophysics Source Code Library, record ascl:1102.026, Feb.,
  2011.

\bibitem{SPTreview2002PhR}
F.~{Bernardeau}, S.~{Colombi}, E.~{Gazta{\~n}aga} and R.~{Scoccimarro},
  \emph{{Large-scale structure of the Universe and cosmological perturbation
  theory}}, \href{https://doi.org/10.1016/S0370-1573(02)00135-7}{\emph{Physics
  Reports} {\bfseries 367} (2002) 1}
  [\href{https://arxiv.org/abs/astro-ph/0112551}{{\ttfamily
  astro-ph/0112551}}].

\bibitem{Paul2023}
S.~Paul, M.G.~Santos, Z.~Chen and L.~Wolz, \emph{{A first detection of neutral
  hydrogen intensity mapping on Mpc scales at $z\approx 0.32$ and $z\approx
  0.44$}}, {\emph{arXiv e-prints} (2023) }
  [\href{https://arxiv.org/abs/arXiv:2301.11943}{{\ttfamily
  arXiv:2301.11943}}].

\end{thebibliography}\endgroup






\end{document}